\documentclass{optica-article}

\journal{opticajournal} %

\articletype{Research Article}

\usepackage{graphicx}%
\usepackage{graphicx}%
\usepackage{multirow}%
\usepackage{siunitx}
\usepackage{amsmath, amsfonts}%
\usepackage{mathrsfs}%
\usepackage[title]{appendix}%
\usepackage{xcolor}%
\usepackage{textcomp}%
\usepackage{manyfoot}%
\usepackage{booktabs}%
\usepackage{algorithm}%
\usepackage{algorithmicx}%
\usepackage{algpseudocode}%
\usepackage{listings}%
\usepackage{lineno}

\begin{document}

\title{ConvRML: high-quality lensless imaging with random multi-focal lenslets}

\author{Leyla A. Kabuli,\authormark{1,*} Clara S. Hung,\authormark{1} Vasilisa Ponomarenko,\authormark{1} Eric Markley,\authormark{1}, and Laura Waller\authormark{1}}

\address{\authormark{1}Department of Electrical Engineering and Computer Sciences, University of California, Berkeley, California 94720, USA}

\email{\authormark{*}lakabuli@berkeley.edu} %

\begin{abstract*} 
Mask-based lensless imagers use simple optics and computational reconstruction to design compact form factor cameras with compressive imaging ability. However, these imagers generally suffer from poor reconstruction quality. Here, we describe several advances in both hardware and software that result in improved lensless imaging quality. First, we use a precision-manufactured random multi-focal lenslet (RML) phase mask to produce improved measurements with reduced multiplexing. Next, we implement a ConvNeXt-based reconstruction architecture, which provides up to \SI{4.6}{\dB} improvement in peak signal-to-noise ratio over state-of-the-art attention-based architectures. Finally, we establish a parallel imaging setup that simultaneously images a scene with RML, diffuser, and lens systems, with which we collect datasets with 100,000 measurements for each system, to be used for reconstruction model training and evaluation. Using this standardized system, we quantify the improved measurement quality of the RML compared to a diffuser using the modulation transfer function and mutual information. Our ConvRML system benefits from both the optical and the computational developments presented in this work, and our contributions establish resources to support the continued development of high-quality, compact, and compressive lensless imagers.

\end{abstract*}

\section{Introduction}
Computational imaging systems jointly design optical hardware and computational algorithms to push the limits of image capture and interpretation. In mask-based lensless imaging, a traditional lens is replaced by a thin amplitude or phase mask that patterns incoming light into a multiplexed sensor measurement, which is then computationally decoded to reconstruct an image.  This two-part design offers a flexible alternative to conventional optics, enabling compact, low-cost, and single-shot systems that support compressive encoding and recovery, including three-dimensional (3D)~\cite{DiffuserCam, PhlatCam, xueCM2, Feng2022Learned} and extended field-of-view (FOV)~\cite{YaokevineFOV, LenslessInfoTheory, claraExtendedFOV, KCextendedSVAview} imaging. With these capabilities, lensless imaging systems are promising for applications ranging from photography~\cite{
DiffuserCam, PhlatCam}, machine vision~\cite{Tan2019FlatCamFace}, and display calibration~\cite{AksitDisplayCalibration} to \textit{in vivo} microscopy~\cite{GraceDiffuserScope, Feng2025DeepInMiniscope} and mesoscopy~\cite{xueCM2, biomesoscope, coreyfundus}. 

A central challenge in lensless imaging is image quality. Ill-posed inverse problems, model mismatch, and compressive encoding all result in lower reconstruction quality than traditional lens systems. Although reconstruction algorithms with increased computational power can partially compensate, their performance is fundamentally limited by the amount of information encoded in the measurements~\cite{LenslessInfoTheory}. Improving lensless image quality therefore requires developing both optical encoders that produce better measurements \textit{and} algorithms that can utilize those measurements. 

A primary research direction in lensless imaging has focused on optical encoder design to improve measurement quality. Phase masks have gained widespread use due to their light efficiency compared to amplitude masks~\cite{asif2017flatcam}.
Initial phase mask-based systems used off-the-shelf diffusers~\cite{DiffuserCam} or heuristically-designed  masks~\cite{PhlatCam} that produce large distributed point spread functions (PSFs). We refer to such PSFs that distribute light over a large area as \textit{high multiplexing}, since they map each point in the scene to many pixels on the sensor, resulting in a high-multiplexing measurement. Multiplexing is necessary for compressed sensing and often useful because of its redundancy, but for dense natural scenes, high multiplexing levels reduce measurement contrast, limiting the image quality of the system.

As a result, optical designs have shifted toward using less multiplexing; for example, regular microlens arrays~\cite{xueCM2, Feng2021Geomscope, YaokevineFOV} and random lenslet arrays~\cite{Feng2022Learned, Feng2025DeepInMiniscope, NickRollingShutter, GraceDiffuserScope, LenslessInfoTheory} produce PSFs that span a large area but have only a few sparse points. Implementable designs have been constrained by fabrication limitations, with previous systems relying on hand-assembled~\cite{Feng2025DeepInMiniscope} or manually-fabricated optical elements~\cite{NickRollingShutter, LindaFourierDiffuserScope, GraceDiffuserScope}, which limit precision in surface control and reproducibility. As precision fabrication of freeform optical elements becomes increasingly accessible, lenslet-based designs offer a clear path to improved lensless measurement quality.

A complementary direction has focused on developing reconstruction algorithms to better utilize multiplexed measurements. Decoders have shifted from iterative algorithms with physics-based constraints~\cite{ADMM, FISTA} to hybrid~\cite{MonakhovaLearning, cai2024phocolens, kingshott2022unrolled} and pure deep learning-based approaches~\cite{UNet, PanTransformer} with increasing computational and architectural complexity. This includes attention-based architectures designed to explicitly capture long-range dependencies introduced by multiplexing PSFs~\cite{PanTransformer}. Much of this algorithmic development has progressed independently of optical design, with new architectures typically evaluated on existing open-source datasets~\cite{MonakhovaLearning,Tan2019FlatCamFace, FlatNet, PanTransformer}. However, these datasets consist of measurements taken with high-multiplexing PSFs, rather than newer, lower-multiplexing lenslet-based designs. As a result, it remains unknown which architectures can best utilize higher-quality measurements, and how much reconstructions can improve when algorithms are paired with better optical encoders.

\begin{figure}[t]
    \includegraphics[width=\textwidth]{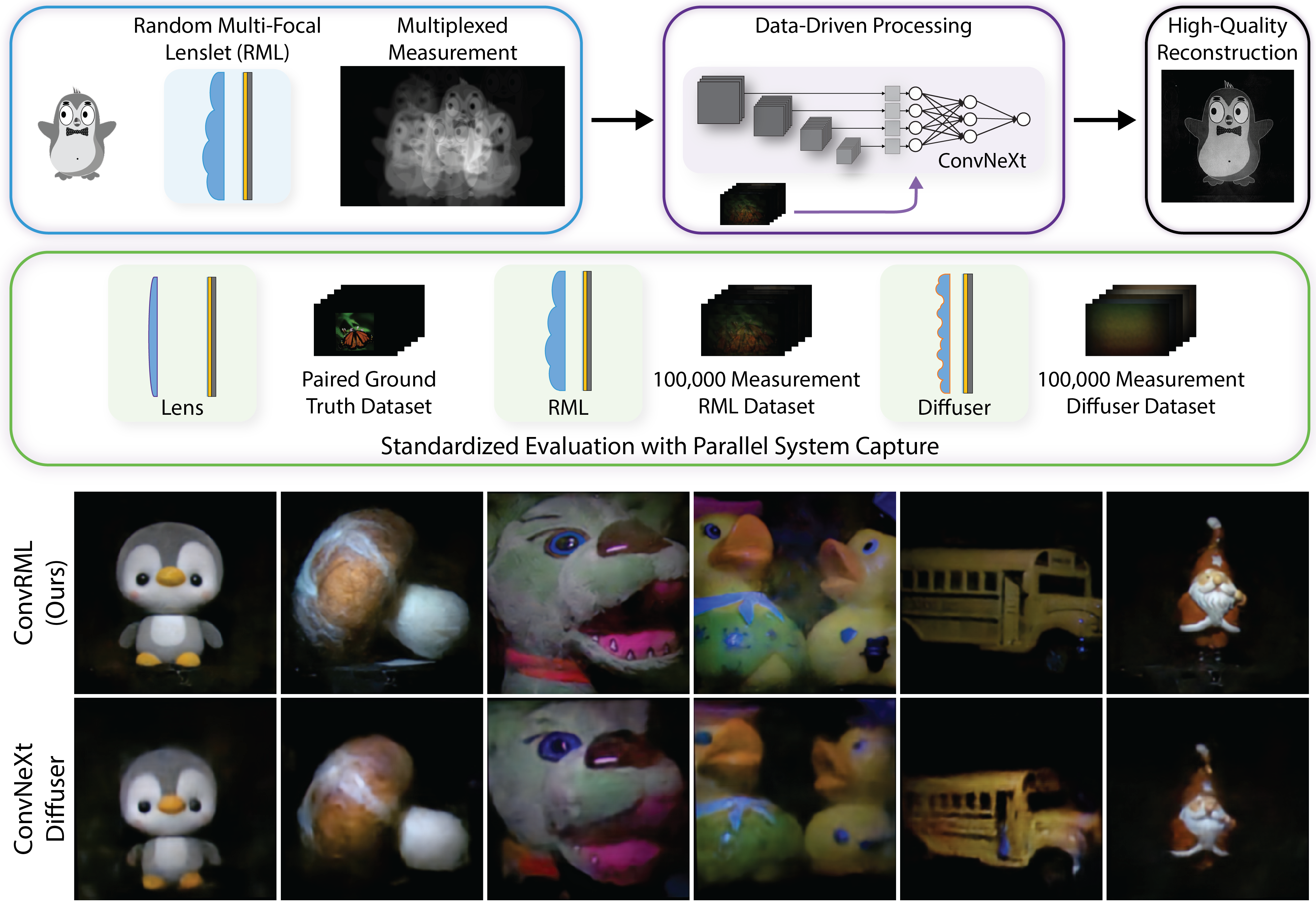}
    \centering
    \caption{\textbf{ConvRML system overview.} (top) Our high-quality lensless imaging system combines a random multi-focal lenslet (RML) phase mask with a ConvNeXt reconstruction architecture. (middle) To facilitate data-driven training of reconstruction architectures and controlled system comparisons between a lens, a diffuser, and our RML, we collect and provide open-source datasets with 100,000 measurements for each system.
    (bottom) ConvNeXt models trained on displayed image datasets for the RML and diffuser lensless imagers produce high-quality reconstructions of real-world objects.}
    \label{fig:intro}
\end{figure}

In this work, we present ConvRML, a lensless imaging system that pairs precision-manufactured optical elements with modern convolutional neural networks and standardized training datasets to advance high-quality lensless imaging (Fig.~\ref{fig:intro}).
We show that a random multi-focal lenslet (RML) phase mask improves measurement quality and can be precisely fabricated using recent advances in optical manufacturing. Our deep learning-based reconstruction uses a ConvNeXt architecture, 
which achieves up to \SI{4.6}{\dB} peak signal-to-noise ratio (PSNR) improvement over state-of-the-art attention-based architectures, successfully processing long-range dependencies in multiplexed measurements even with limited training data.

To facilitate comprehensive evaluation and continued system development, we contribute multiple datasets each with 100,000 measurements, captured using a standardized parallel hardware setup 
that enables controlled comparison across optical encoders and reconstruction algorithms. Using this setup, we quantify the performance improvements of ConvRML relative to a diffuser phase mask in both optical encoding and image reconstruction, and present high-quality reconstructions for both systems that generalize to real-world objects.
Together, our work demonstrates that improved measurement quality paired with powerful algorithms facilitates high-quality lensless imaging reconstructions, and establishes resources for continued lensless imaging system development. 

\section{Results}\label{sec2}

Lenses collect and focus light such that a point in the scene is imaged to a focused point on the sensor. For a fixed numerical aperture (NA), reducing the focal length of the lens produces a more compact system but proportionally reduces the FOV (Fig.~\ref{fig:analysis}a). Achieving both a compact system and a wide FOV requires high-NA optics, which are typically thick, expensive and prone to aberrations. In contrast, lensless imagers exchange the lens for a thin multiplexing element placed close to the sensor, maintaining both a compact form factor and a wide FOV (Fig.~\ref{fig:analysis}a). As a consequence, points in the scene must be imaged onto multiple pixels of the sensor; this multiplexing effect is why lensless imagers require a decoding algorithm for image reconstruction. 

Each imaging system can be characterized by its PSF. With no multiplexing, a conventional lens system produces a point PSF, whose width is determined by the NA of the system.
Gaussian diffusers~\cite{DiffuserCam} instead refract light using bumpy surfaces, producing a high-multiplexing caustic pattern (Fig.~\ref{fig:analysis}b). Lenslet-based masks use arrays of lenslets that produce a lower-multiplexing, high-contrast PSF consisting of multiple focal spots with minimal diffuse background.

For photography-scale lensless imaging systems, %
the measurement formation model can be approximated by a shift-invariant convolution,
$\mathbf{y} %
   \approx 
    \mathcal{N} (\text{crop} (\mathbf{x} \ast \mathrm{h}))$.
The object $\mathbf{x}$ is convolved with the system PSF $\mathrm{h}$, cropped by the sensor extent, and corrupted by detection noise $\mathcal{N}(\cdot)$, typically modeled as signal-independent read noise or signal-dependent shot noise~\cite{Kabuli2025NoiseModel, Liu2025LowLightNoise}. This approximate forward model is useful for system analysis, but becomes less accurate with shorter working distances or  larger FOVs~\cite{DiffuserCam}, where off-axis light can result in spatially varying PSFs~\cite{SVFourierNet} that require more complicated forward models~\cite{GraceDiffuserScope, MultiWienerNet}. Our deep learning-based reconstruction methods do not explicitly assume spatial invariance or impose a specific forward model, so they may be flexible enough to implicitly learn spatially varying effects (Supplement Fig.~\ref{fig:svpsfmtf}).

\subsection{Random multi-focal lenslet phase mask}

Our lensless imaging system uses a random multi-focal lenslet (RML) phase mask as the optical encoder, which consists of overlapping lenslets with a range of focal lengths placed at pseudo-random positions across the aperture. %
Compared to a regular lenslet array, pseudo-random lenslet positions minimize correlations between shifted copies of the system PSF, which reduces ambiguity during reconstruction~\cite{GraceDiffuserScope, LindaFourierDiffuserScope, KabuliReplica}. 
Limiting the number of lenslets reduces multiplexing to improve measurement quality for dense natural scenes, and high fill factor across the mask aperture minimizes stray light leakage. 
With recent advances in high-resolution, large-area two-photon lithography~\cite{nanoscribefabex}, these designs can now be precisely, quickly, and reproducibly fabricated. 

Because our phase mask has lenslets focused at various depths, the imager maintains sharp PSF features even with variations in the sensor-to-mask distance. Compared to caustic masks or uni-focal lenslet arrays, which should be precisely positioned at a specific distance for sharp PSF features, this reduces our design's sensitivity to system alignment~\cite{Kabuli2023RML}. %

\begin{figure}
\centering
    \includegraphics[width=\textwidth]{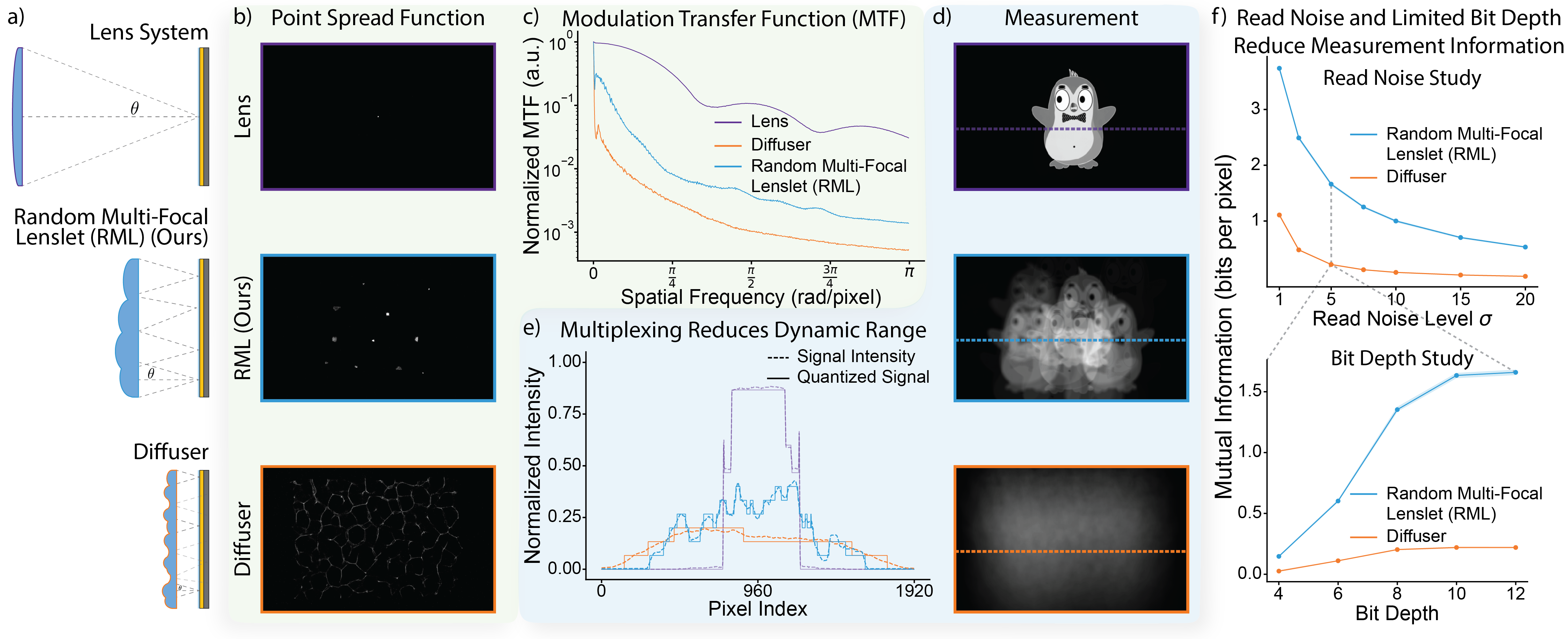}
    \caption{\textbf{Random multi-focal lenslet (RML) phase masks result in good measurement quality.} 
    a) Illustration of different optical encoders with a fixed numerical aperture $\theta$ and b) experimentally-measured point spread functions (PSFs). A lens focuses light to a single focal point, with the field of view (FOV) decreasing for shorter focal lengths. The RML distributes light across a small number of focal points, preserving a wide FOV with minimal diffuse background. The diffuser both focuses and scatters light, resulting in a high-multiplexing caustic pattern. PSFs are contrast stretched for visualization.
    c) Modulation transfer functions (MTFs) for each design, with the lens system serving as an ideal reference. The RML has improved frequency transfer compared to the diffuser, especially at high spatial frequencies. 
    d) Simulated measurements for each imaging system. The high-multiplexing diffuser PSF generally results in lower contrast measurements, while RML measurements preserve larger local intensity variations.
    e) Cross-sections of each measurement along with dashed lines showing the loss of dynamic range with quantization.
    f) Mutual information evaluation of the effects of read noise and limited bit depth. The RML maintains higher information than the diffuser as noise increases (top) and retains nonzero information at low bit depths while the diffuser has approximately zero information (bottom).
    }
    \label{fig:analysis}
\end{figure}

Direct comparison to previously-proposed lensless imagers is challenging because of variations in system implementation. 
Differences in sensor specifications, system magnification, compactness, and illumination conditions all affect measurement quality, confounding optical encoder comparisons. To isolate the effect of the optical encoder system, we establish a parallel imaging setup in which designs are evaluated under identical imaging conditions with the same sensor, illumination, object magnification, and careful alignment (Fig.~\ref{fig:rml_vs_dc_evaluation}a). In our standardized evaluation, we compare our RML to a diffuser and a lens, which serves as a reference for target image quality without multiplexing.

Using the experimentally captured PSFs shown in Fig.~\ref{fig:analysis}b, we calculate the modulation transfer function (MTF) for each system to evaluate the frequency transfer characteristics of the optical encoder (Fig.~\ref{fig:analysis}c). As expected, both the RML and diffuser have worse frequency transfer than the lens due to their multiplexed PSFs. The RML has higher frequency transfer than the diffuser across all spatial frequencies due to the reduced multiplexing and sharp focal spots in its PSF. These frequency transfer characteristics are maintained at off-axis positions in the FOV (Supplement Fig.~\ref{fig:svpsfmtf}). The RML similarly outperforms PhlatCam (Supplement Fig.~\ref{fig:phlatcamMTF}), although we note that the PhlatCam PSF was captured under different experimental imaging conditions~\cite{PhlatCam}.

High-multiplexing encoders generally reduce measurement contrast, requiring costly sensors with low noise and high bit depth~\cite{Li2025LenslessSensor} to capture small intensity variations that can otherwise be obscured by noise or lost during quantization. Comparing simulated measurements for each system (Fig.~\ref{fig:analysis}d), the RML generates shifted object copies that preserve fine details while the diffuser produces a measurement that has no visually interpretable object structure.
To illustrate, we plot measurement cross-sections in Fig.~\ref{fig:analysis}e. The diffuser has minimal signal variation, whereas the RML produces high-contrast signals and the lens has sharp transitions corresponding to object edges. To indicate the effects of limited sensor bit depth, we overlay a 4-bit quantization of the signals (dashed lines, Fig.~\ref{fig:analysis}e). With heavy quantization, most signal variations for the diffuser fall within a single quantization bin and are effectively lost, whereas the RML maintains both distinguishable features and multiplexing ability.  %

Finally, we evaluate measurement quality using mutual information (MI), which quantifies how much object information is available in measurements, capturing the combined effects of optical encoding, object statistics, and measurement noise~\cite{infotheorypaper, LenslessInfoTheory}. We first study the effect of read noise by calculating MI for simulated measurements of natural images~\cite{mirflickr1M} with increasing standard deviations $\sigma$ of read noise (Fig.~\ref{fig:analysis}f, top). Across all noise levels, the RML encodes more information than the diffuser for natural images. Then, we evaluate the effect of quantization noise resulting from limited bit depth. For a fixed read noise level ($\sigma = 5$), we quantize measurements to effective bit depths of 4 to 12 bits and calculate MI in each case (Fig.~\ref{fig:analysis}f, bottom), finding that the RML continues to outperform the diffuser across all bit depths. At 4 bits, the diffuser retains almost no information, while the RML measurements continue to encode object information. This result is consistent with the dynamic range visualization in Fig.~\ref{fig:analysis}e, where most diffuser signal variations are lost to quantization. Our evaluations complement prior MI analysis of experimental measurements, which showed that the RML encodes more information than the diffuser for natural images~\cite{LenslessInfoTheory}.

\subsection{100,000 measurement Parallel Lensless Dataset} 

Modern data-driven reconstruction architectures, such as attention-based architectures or large convolutional neural networks, require large training datasets. For example, recent transformer-based models were trained with 60,000 lensless measurements~\cite{PanTransformer}. However, available experimental datasets for phase mask-based systems have only 10,000 measurements for PhlatCam~\cite{FlatNet} and 25,000 measurements for DiffuserCam~\cite{MonakhovaLearning}, with no open-source datasets available for lenslet-based systems. 

Here, we introduce an open-source experimental dataset of 100,000 measurements for multiple imaging systems: our RML lensless imager, a diffuser lensless imager, and a conventional lens system (which we treat as ground truth). 
To automate parallel data acquisition, we use synchronized hardware and software control (Fig.~\ref{fig:rml_vs_dc_evaluation}a). A photograph of the experimental hardware setup is provided in Supplement Fig.~\ref{fig:dataset}.
Each imaging system uses the same sensor, illumination, and object magnification to ensure that differences in system performance are due to the optical encoder system rather than experimental variations. 
Together, our Parallel Lensless Dataset (PLD) provides the largest open-source experimental lensless imaging datasets to date. We use these datasets to provide a standardized system-level comparison between imagers, and hope that they serve as resources to facilitate future development of reconstruction architectures. 

\subsection{ConvRML system evaluation} 

We now evaluate our complete ConvRML imaging system, incorporating improved optical measurements from the RML, a modern ConvNeXt architecture for reconstruction~\cite{ConvNeXtOriginal}, and our open-source dataset.

\begin{figure}
\centering
    \includegraphics[width=\textwidth]{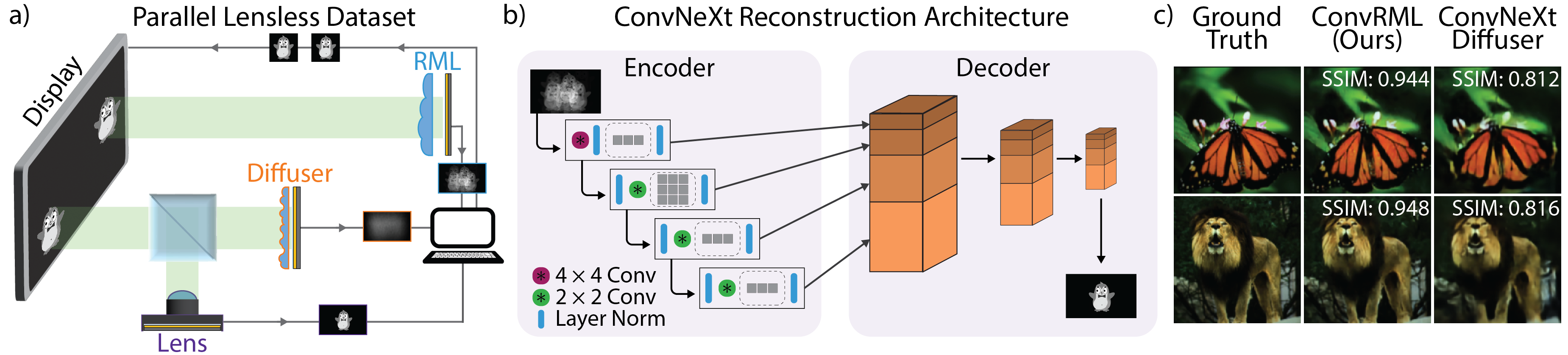}
    \caption{\textbf{Standardized experimental comparison of RML and diffuser systems.} 
    a) Parallel dataset acquisition with RML, diffuser, and lens (ground truth) systems under identical imaging conditions facilitates standardized system-level comparisons.
    b) The ConvNeXt-based reconstruction architecture creates multi-scale feature maps, which are concatenated and progressively compressed. 
    c) Example reconstructions comparing ground truth, our RML with ConvNeXt reconstruction (ConvRML), and a diffuser with ConvNeXt reconstruction. Using an RML achieves higher reconstruction quality than the diffuser, as quantified by the structural similarity index measure (SSIM). 
    }
    \label{fig:rml_vs_dc_evaluation}
\end{figure}

Lensless image reconstruction is challenging in part due to the spatial extent of system PSFs, which produce multiplexed measurements with long-range dependencies.
ConvNeXt is a convolutional neural network architecture that achieves large effective receptive fields through cascaded convolutional layers while maintaining moderate computational requirements~\cite{ConvNeXtOriginal, LargeCNNKernels}. 
We expect this neural network, with large receptive fields capable of processing long-range dependencies from multiplexing PSFs, to be suitable for lensless image reconstruction. 
Our reconstruction architecture combines the ConvNeXt encoder, which consists of a series of downsampling blocks, with concatenation and compression operations in the decoder (Fig.~\ref{fig:rml_vs_dc_evaluation}b).

We train separate ConvNeXt models for the RML and diffuser lensless imagers, using 50,000 measurements from the PLD for each system. By keeping all training procedures and dataset sizes consistent, we isolate the interaction between optical encoding and reconstruction architecture. 

Figure~\ref{fig:rml_vs_dc_evaluation}c shows reconstruction results using our ConvNeXt reconstruction models with both RML and diffuser systems. ConvRML reconstructions better recover fine spatial detail, contrast, and color fidelity, with test set improvements of \SI{7.014}{\dB} PSNR and 0.103 structural similarity index measure (SSIM) relative to the diffuser system. Higher-quality RML reconstructions are maintained when using an iterative physics-based reconstruction algorithm~\cite{ADMM} (Supplement Fig.~\ref{fig:admmcomparison}), which confirms that the reconstruction improvement from the RML is not specific to the ConvNeXt model.
Since these comparisons control for algorithmic variability, dataset size, and experimental conditions, we can conclude that the RML measurements directly translate to higher-quality reconstructions than with the diffuser system.

\subsection{Comparison of reconstruction architectures}

Next, we isolate the effects of reconstruction architecture and training data. We compare multiple modern deep learning architectures, including attention-based architectures,
on both an existing benchmark dataset as well as our contributed datasets, in order to determine how dataset size and multiplexing levels influence the performance of different architectures. 

For comparison to our ConvNeXt architecture, we provide lensless imaging implementations of several attention-based architectures. 
First, we contribute a lightweight Vision Transformer~\cite{ViT}. This small attention-based architecture transforms the input into a series of flattened image patches, which are processed with global self-attention layers to capture long-range dependencies. Second, we contribute a Swin Transformer~\cite{SwinTransformer}, a large-capacity architecture that utilizes shifted-window attention for a hierarchical representation that captures both local and global features. 
We also compare to the Lensless Imaging Transformer~\cite{PanTransformer}, which is a moderate-capacity attention-based reconstruction architecture designed for lensless imaging. 
As a deep learning reconstruction baseline, we include the standard lensless U-Net architecture~\cite{MonakhovaLearning}.

We first compare reconstruction architectures by training all models on the DiffuserCam Lensless Mirflickr Dataset (DLMD)~\cite{MonakhovaLearning}, which consists of 25,000 measurements from a high-multiplexing diffuser. This widely-used benchmark for data-driven lensless imaging reconstruction provides a standardized reference for direct model comparison. 

\begin{figure}[t]
\centering
    \includegraphics[width=\textwidth]{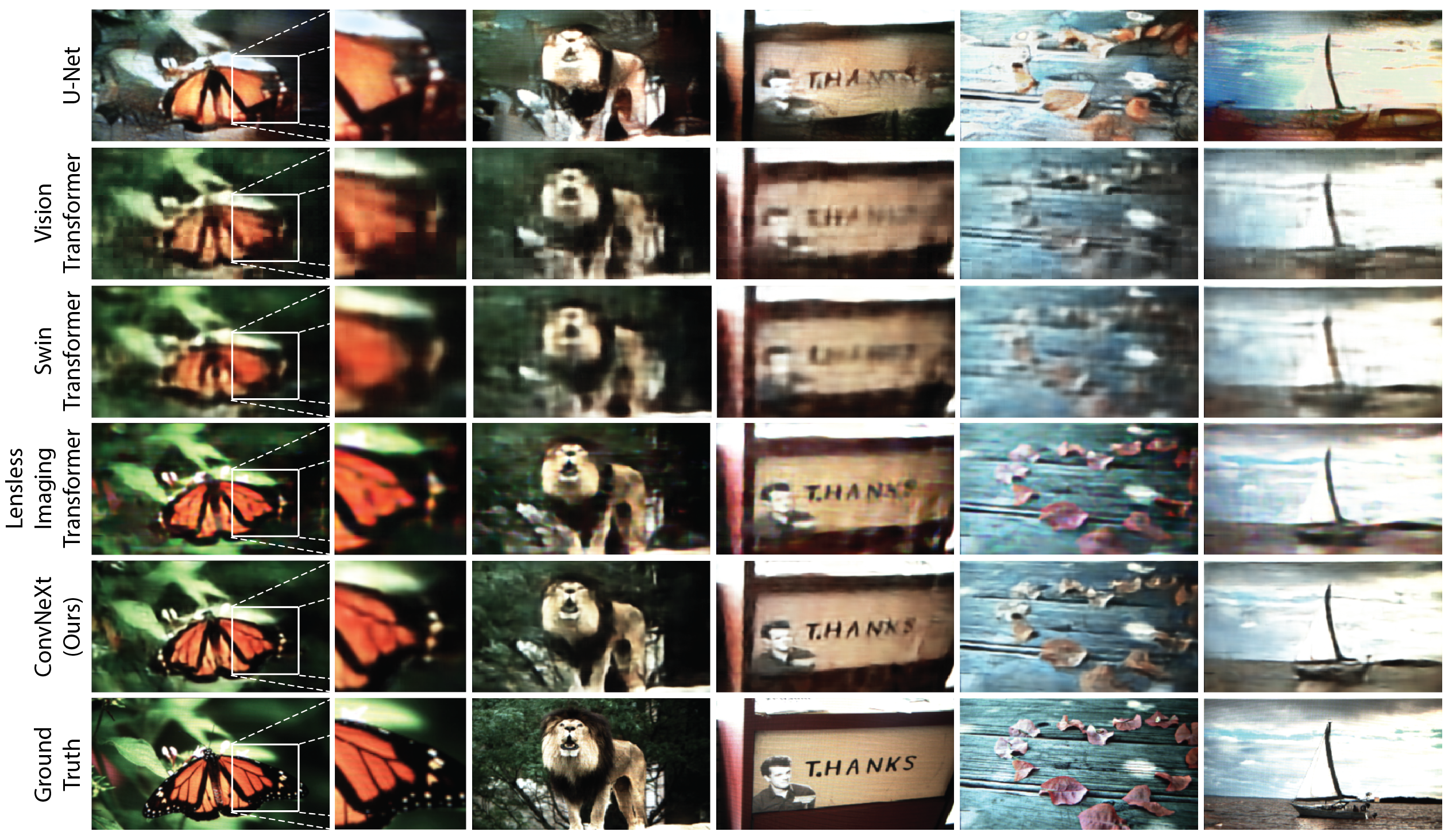}
    \caption{\textbf{Comparison of reconstruction models evaluated on the DiffuserCam Lensless Mirflickr Dataset~\cite{MonakhovaLearning}.} Our ConvNeXt model provides reconstructions that most closely match ground truth images for this high-multiplexing diffuser system. Corresponding quantitative reconstruction metrics are reported in Table~\ref{tab:mirflickr_comparison}.}
    \label{fig:dlmd_evaluation}
\end{figure}

We evaluate our ConvNeXt, the Lensless Imaging Transformer~\cite{PanTransformer}, the Swin Transformer, and the Vision Transformer against the U-Net baseline reported for the DLMD~\cite{MonakhovaLearning}. Reconstruction results are shown in Fig.~\ref{fig:dlmd_evaluation}. Compared to the ground truth images (bottom row), ConvNeXt has the highest reconstruction quality, with improved color fidelity and recovery of fine image details. The compact Vision Transformer and the large Swin Transformer both exhibit visible blocking artifacts and blurred image features that are qualitatively worse than the U-Net. The Lensless Imaging Transformer provides improved results over the U-Net, especially in color fidelity, but has worse resolution than ConvNeXt. 

\begin{table}[h]
\centering
\caption{Reconstruction quality and parameter count for models evaluated on the DiffuserCam Lensless Mirflickr Dataset~\cite{MonakhovaLearning}.}
\label{tab:mirflickr_comparison}
\begin{tabular}{lccccc}
\toprule
\textbf{Model}
& \textbf{PSNR $\uparrow$}
& \textbf{SSIM $\uparrow$}
& \textbf{LPIPS $\downarrow$}
& \textbf{Parameters} \\
\midrule
U-Net %
& 17.293 & 0.583 & 0.375 & 11.8M \\
Vision Transformer    
& 21.256 & 0.628  & 0.421 & 1.0M \\
Swin Transformer 
&  21.192 & 0.633  & 0.483 & 59.4M\\
Lensless Imaging Transformer%
& 22.468 & 0.673  & 0.393 & 16.1M\\
ConvNeXt (Ours)       
&  \textbf{25.369} & \textbf{0.762}  & \textbf{0.279} & 36.0M \\

\bottomrule
\end{tabular}
\end{table}

Quantitative reconstruction metrics are reported in Table~\ref{tab:mirflickr_comparison}. ConvNeXt achieves the highest PSNR and SSIM and lowest learned perceptual image patch similarity (LPIPS)~\cite{LPIPS} among all evaluated models while maintaining moderate model capacity, with fewer parameters than the Swin Transformer. These results demonstrate that a carefully-designed convolutional architecture can outperform attention-based architectures on a high-multiplexing system with a moderately-sized dataset.

Next, we compare reconstruction architectures for our lower-multiplexing RML using a larger measurement dataset. Although ConvNeXt outperformed attention-based models on the DLMD benchmark, it is possible that the 25,000 measurement DLMD dataset may have been insufficient for training the attention-based architectures. Therefore, we investigate whether increasing dataset size changes the relative performance of these models for RML measurements. 

We study tradeoffs between model capacity, dataset size, and training cost by training the ConvNeXt model on subsets ranging from 25,000 to 100,000 RML measurements from the PLD (Supplement Fig.~\ref{fig:dataset_size_sweep}). Reconstruction quality improves with dataset size and plateaus at 75,000 measurements, while training time increases substantially. We use a dataset size of 50,000 measurements for all subsequent evaluations to increase dataset size while keeping training time tractable across all architectures. 

\begin{figure}[hbtp]
\centering
    \includegraphics[width=0.75\textwidth]{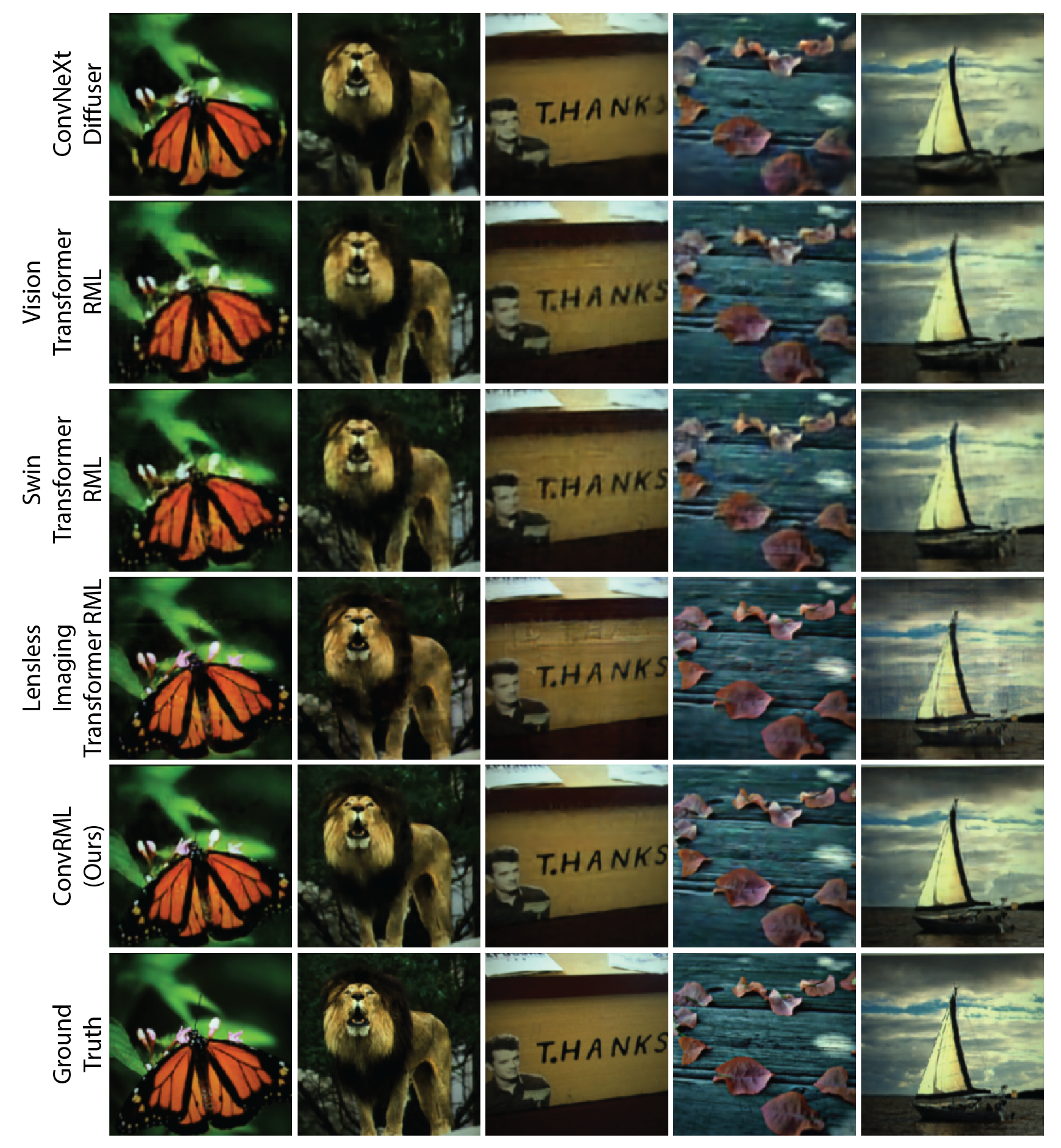}
    \caption{\textbf{Comparison of reconstruction models evaluated on our Parallel Lensless Dataset.} Our ConvRML system provides reconstructions that are best matched to ground truth images for the low-multiplexing RML dataset. ConvNeXt reconstructions for high-multiplexing diffuser measurements are provided as a reference. Corresponding quantitative reconstruction metrics are reported in Table~\ref{tab:rml_comparison}. Both the RML and the ConvNeXt reconstruction architecture contribute to image quality improvements over state-of-the-art.}
    \label{fig:50k_algorithms}
\end{figure}

Using this dataset size, we compare ConvNeXt to the Swin Transformer, the Lensless Imaging Transformer, and the Vision Transformer for RML measurement reconstruction, as well as ConvNeXt reconstructions of diffuser measurements as a baseline.
Reconstruction examples are shown in Fig.~\ref{fig:50k_algorithms} and quantitative reconstruction metrics are reported in Table~\ref{tab:rml_comparison}.
All models benefit from the larger training dataset, although the compact Vision Transformer still suffers from patching artifacts as previously seen in Fig.~\ref{fig:dlmd_evaluation}. Among the three attention-based models, the Lensless Imaging Transformer best recovers fine details and perceptual features, as reflected by its lower LPIPS value, but image artifacts result in reduced PSNR and SSIM.
ConvNeXt consistently achieves the highest reconstruction quality.

\begin{table}[hbtp]
\centering
\caption{Reconstruction quality for models evaluated on 50,000 measurements from the Parallel Lensless Dataset.}
\label{tab:rml_comparison}
\begin{tabular}{lccc} %
\toprule
\textbf{Model}
& \textbf{PSNR $\uparrow$}
& \textbf{SSIM $\uparrow$}
& \textbf{LPIPS $\downarrow$} \\
\midrule
ConvNeXt Diffuser & 30.245 & 0.838 & 0.170 \\ %
Vision Transformer RML  & 33.349 & 0.880 & 0.160 \\ %
Swin Transformer RML & 33.461 & 0.882 & 0.148\\ %
Lensless Imaging Transformer RML  & 32.625 & 0.870 & 0.119  \\ %
ConvRML (Ours)       & \textbf{37.259} & \textbf{0.941} & \textbf{0.0604}   \\ %

\bottomrule
\end{tabular}
\end{table}

These studies demonstrate that attention-based architectures with increased model capacity and architectural complexity are not strictly necessary to achieve high-quality lensless image reconstruction, especially when reduced multiplexing improves measurement quality. 
Our ConvNeXt reconstruction generalizes across both low and moderate data regimes, and both high- and low-multiplexing encoders, validating its suitability as a flexible reconstruction architecture for lensless imaging.

\subsection{Experimental generalization to real-world objects}

In previous sections, we compared optical designs and reconstruction algorithms using natural images displayed on a monitor under controlled illumination. This setting facilitated large-scale data acquisition and standardized comparison. Here, we evaluate models' generalization ability by reconstructing physical objects placed in front of the lensless imaging systems. These real-world imaging conditions introduce additional sources of variability that can present significant challenges for algorithm generalization, including object-dependent signal intensities, variations in object positions and depths, system non-idealities, and uncontrolled illumination~\cite{BezzamIllumination}. %

Objects are hand-aligned for approximately equal positioning in front of each lensless imager, with inevitable variations in alignment, magnification, and FOV. 
While the display emits light uniformly across the object FOV, real objects require external illumination, with captured measurements relying on light reflected from object surfaces. Object specularity affects measurement quality, with matte objects requiring stronger and closer illumination than specular objects, and reflective surfaces introducing bright highlights.

Using ConvNeXt models trained on our datasets of displayed natural images, we reconstruct real objects for both RML and diffuser systems. Figure~\ref{fig:intro} shows reconstruction results for objects with varying materials and surface properties, including matte fabric, plastic, metal, and painted wood. Additional examples are provided in Supplement Fig.~\ref{fig:supp_experimental}. Reconstructions for both systems are high quality, and the differences in resolution and image quality between the RML and diffuser system reconstructions are consistent with previous sections. The RML successfully recovers fine textures of objects with matte surfaces, including the fibers on the mushroom, the felt on the penguin, and the stitching on the dog, whereas the diffuser struggles to recover these details. Sharp edges and patterns on the plastic ducks are recovered by the RML but missed on the diffuser. The metal school bus highlights reconstruction challenges from specular effects and variable illumination, which cause distortion in the diffuser but not the RML.

\section{Discussion}\label{sec3}
ConvRML pairs a precision-manufactured random multi-focal lenslet (RML) phase mask with a ConvNeXt convolutional neural network reconstruction architecture. Our phase mask produces 
high-quality measurements that preserve measurement contrast and exhibit tolerance to sensor read noise and quantization error. Our ConvNeXt reconstruction outperforms attention-based models and provides high-quality image reconstructions that generalize to real objects. Throughout this work, we standardize comparisons between lensless imaging systems with parallel data acquisition, and provide datasets of 100,000 measurements for both the RML and diffuser lensless imagers, as well as a lens system for comparison. Together, our contributions showcase image quality improvements made possible by combining advances in optical encoder design with data-driven reconstruction architectures.

The RML used in this work is a general-purpose heuristic design that supports both lensless imaging and Fourier-plane modulation~\cite{LindaFourierDiffuserScope}. It therefore prioritizes reduced multiplexing over extreme compactness, while still maintaining sensor-to-mask distances comparable to existing lenslet-based lensless imagers~\cite{Feng2022Learned}. Our RML is distinguished from previous lenslet-based systems~\cite{LindaFourierDiffuserScope, GraceDiffuserScope, Feng2025DeepInMiniscope, xueCM2, NickRollingShutter, Feng2021Geomscope, Feng2022Learned} by the use of randomly positioned multi-focal lenslets for photography-scale imaging, and precise, reproducible fabrication of our optical element. As the focal length distribution of the RML should provide an extended depth of focus (sensor-to-mask distance tolerance) at \SI{}{\cm}-scale working distances, we focus on 2D imaging with data-driven reconstruction. Multi-focal lenslets are also naturally suited for depth-varying imaging, and the RML can be redesigned for ultra-compact, 3D, or extended FOV imaging by adjusting focal lengths and lenslet distributions. Future work can also optimize RML designs to maximize task performance and measurement quality using end-to-end~\cite{loene2e, originale2evincent} or information-driven techniques~\cite{LenslessInfoTheory, idealio}.

Our reconstruction approach advances pure deep learning models, which reconstruct images without imposing explicit physics-based constraints such as a calibrated system PSF. Within this scope, our ConvNeXt architecture demonstrates better performance than attention-based architectures while maintaining moderate data and computational requirements. We use a 50,000 measurement subset of the PLD as a practical comparison setting, but large architectures, such as the Swin Transformer, may benefit from additional training measurements. Future work can continue to explore tradeoffs between model capacity, dataset size, and practical training considerations. 
Incorporating diffusion-based enhancement~\cite{cai2024phocolens} or hybrid two-step approaches~\cite{MonakhovaLearning, DeepLIR}, which require calibrated PSFs and system models,
could further improve reconstruction quality.
Finally, although the systems studied here are approximately shift-invariant, exploring ConvNeXt-based reconstruction for spatially varying systems and incorporating existing spatially varying architectures~\cite{xue2022CM2V2, KCextendedSVAview, SVFourierNet} are promising directions, especially for \textit{in vivo} applications.

A unique contribution of our work is a parallel data acquisition system with open-source 100,000 measurement datasets for both the RML and diffuser lensless imagers. Existing open-source datasets are substantially smaller~\cite{MonakhovaLearning, BezzamIllumination, BezzamGeneralization, claraPWdataset, PanTransformer} 
and limited to high-multiplexing encoders. We provide the largest open-source lensless datasets available to date, the first open-source dataset for lenslet-based systems, and the first dataset to support controlled evaluation of multiple systems within a single acquisition framework, while also demonstrating that high-quality lensless imaging can be achieved using low-cost, off-the-shelf sensors.
Our contribution provides shared resources for continued work, supporting development of data-driven reconstruction architectures and enabling standardized evaluation of lensless imaging systems.

\section{Materials and methods}\label{sec4}

\paragraph{Experimental setup}
All imaging systems use identical board-level sensors (Basler daA1920-160uc, Sony IMX392). The sensor-to-mask distance is \SI{16}{\mm} for the RML and \SI{4.5}{\mm} for the diffuser. The lens system uses a \SI{6}{\mm} focal length S-mount lens (Evetar M13B0618W, f/1.8), which is incorporated into the optical path using a 50/50 split ratio beamsplitter (ThorLabs BS031) placed between the display and the diffuser~\cite{MonakhovaLearning}. A system enclosure and a divider between the RML and diffuser block stray light (Supplement Fig.~\ref{fig:dataset}).

The RML phase mask is a heuristic design that has pseudo-random lenslet positions to minimize the PSF autocorrelation~\cite{KabuliReplica}. Focal lengths are randomly selected from approximately 13.5~mm to 15.5~mm. %
The phase mask consists of a polydimethylsiloxane (PDMS) replica of a custom-fabricated laser ablation negative mold (PowerPhotonic)~\cite{LindaDissertation}. With updated two-photon lithography techniques, the mask can also be directly fabricated~\cite{nanoscribefabex}. A \SI{3.02}{\mm} by \SI{2.65}{\mm} aperture placed on the RML limits the PSF extent to the sensor region, defining an effective aperture with 11 lenslets (average NA $\approx 0.05$),  %
as visualized on the RML height map (Supplement Fig.~\ref{fig:heightmap}). %
The diffuser system uses a Gaussian diffuser (Luminit $0.5^{\circ}$) with a 4.73 by \SI{3.33}{\mm} aperture.

Systems are manually aligned for approximately equal object magnification, such that reconstructed images have the same object scale across systems. System PSFs are acquired using a \SI{532}{\nm} (Thorlabs CPS 532) laser point source placed at the calibration plane, located \SI{435}{\mm}, \SI{161}{\mm}, and \SI{165}{\mm} from the RML, diffuser, and lens, respectively. A portable monitor (INNOCN 13A1F) is placed at the same plane to display images. For real-world captures, objects are placed at the calibration plane and illuminated with a flashlight (iPhone 14 Pro). Camera exposure, illumination strength, and illumination distance are adjusted for each real-world capture.

\paragraph{Dataset acquisition}
Our 100{,}000-measurement Parallel Lensless Dataset (PLD) and associated system control, fabrication, alignment, and calibration scripts are publicly available~\cite{ConvRML_Repo}.
The PLD uses the first 100,000 images in the MIRFLICKR-1M dataset~\cite{mirflickr1M}, which are displayed on a monitor and captured using an automatic acquisition pipeline. 
Images are cropped to $300\times300$ pixels to control for different image sizes and displayed with a black background. Two identical copies are displayed for each image, with display positions calibrated to center measurements on the corresponding sensors. Image display and capture are synchronized across systems, which allows for scalable, automatic acquisition of 100{,}000${^+}$ image datasets. 

Captured images are $4\times$ downsampled to $300\times480$ pixels to reduce fringe effects~\cite{MonakhovaLearning} and maintain tractable memory requirements for reconstruction. Full-resolution measurements are also provided~\cite{ConvRML_Repo}.
Lens system ground truth images are corrected for distortion and mirrored to match the lensless imagers' orientation (Supplement Fig.~\ref{fig:dataset}). For quantitative evaluation in the same coordinate space, reconstructions from the RML and diffuser systems are computationally aligned to the undistorted ground truth images (Supplement Sec.~\ref{s5}). Images in figures include max normalization for visualization.

\paragraph{System analysis}
Reported MTFs are normalized by their zero frequency (DC) value and radially averaged. Radial spatial frequency is reported in rad/pixel on the sensor pixel grid, where $\pi$ rad/pixel corresponds to the Nyquist frequency. 

Mutual information (MI) estimation~\cite{infotheorypaper} uses the first 10,000 natural images from the MIRFLICKR-1M~\cite{mirflickr1M} dataset as the object distribution. Each $300 \times 300$ pixel cropped image is converted to grayscale and convolved with experimentally measured PSFs. The resulting noiseless measurements are scaled to a maximum intensity of $2^{12}$ photons and downsampled to $50 \times 50$ pixels. Signal-independent Gaussian noise with standard deviation $\sigma = 1.0$ to $\sigma = 20.0$ is added to simulate read noise. Measurements with read noise ($\sigma = 5.0$) are quantized by adding uniform noise corresponding to effective bit depths between 4 and 12 bits. MI is estimated from randomly sampled $32 \times 32$ pixel measurement patches, with 8000 patches used to fit a full Gaussian entropy model and 2000 patches reserved for evaluation.

\paragraph{Reconstruction architectures}

The ConvNeXt reconstruction architecture uses a ConvNeXt-T encoder backbone~\cite{ConvNeXtOriginal} followed by a Feedforward Neural Network (FNN) decoder~\cite{PanTransformer}. The encoder processes $300 \times 480$ pixel inputs through four downsampling stages with stage blocks of (3, 3, 9, 3), producing multi-scale feature representations. 
The decoder aggregates features using bilinear interpolation and channel-wise concatenation to the target $300 \times 480$ pixel image size, 
followed by five $3 \times 3$ convolutional layers with batch normalization and ReLU activations to produce a 3-channel RGB output.

The Swin Transformer~\cite{SwinTransformer} patches the input into $15 \times 15$ pixel patches, uses a window size of 7, a stage depth ratio of (2, 2, 6, 2), an embedding dimension of 128, and attention head counts of (4, 8, 16, 32). 
Multi-scale features are decoded using the same FNN decoder as ConvNeXt. 
The Vision Transformer~\cite{ViT} consists of 6 encoder blocks with 4 attention heads and an embedding dimension of 128, using a $15 \times 15$ pixel patch size and a dropout rate of 0.1. The Vision Transformer decoder consists of four transposed $2 \times 2$ convolutional layers with stride 2, batch normalization, and ReLU activations. 
The Lensless Imaging Transformer is modified to support $300 \times 480$ pixel input and output dimensions, with the remainder of the architecture unchanged from the original implementation~\cite{PanTransformer}. We use a pre-trained U-Net~\cite{MonakhovaLearning} in architecture comparisons on the DLMD dataset (Fig.~\ref{fig:dlmd_evaluation}).

All models are implemented in PyTorch~\cite{pytorch} and trained on a single NVIDIA RTX A6000 GPU, with training times reported in Supplement Table~\ref{tab:pld_train_time} and Table~\ref{tab:dlmd_train_time}. All models use the AdamW optimizer with a mean squared error (MSE) loss and a batch size of 6. All reconstruction mappings are learned directly from paired measurements and ground truth images, with no knowledge of the system PSF or forward model. Input measurements are normalized to $[0, 1]$ prior to training. For evaluation, reconstruction outputs are clamped to $[0, 1]$, and PSNR, SSIM, and LPIPS are averaged over the test set. Inference times are reported in Supplement Table~\ref{tab:inference_times}. Metrics are computed over a cropped reconstruction extent, corresponding to image sizes of $210 \times 380$ pixels and $214 \times 214$ pixels for the DLMD and PLD, respectively.

DLMD dataset evaluation follows the standard data split~\cite{MonakhovaLearning}, reserving the first 1000 images for testing and using the remaining 24,000 images for training. The ConvNeXt model is trained for 35 epochs using a learning rate of $5 \times 10^{-4}$ and weight decay of $10^{-3}$, using a linear warmup followed by cosine annealing. The Swin Transformer follows the same protocol, while the Vision Transformer is trained for 70 epochs to ensure convergence. The Lensless Imaging Transformer is trained for 240,000 steps ($\approx 60$ epochs) using a learning rate of $6 \times 10^{-5}$, weight decay of $10^{-1}$, and the same scheduling process as ConvNeXt.

For the PLD, the first 1000 images are reserved for testing, the subsequent 4000 images for validation, and the remaining images for training. The main text uses a 50,000 measurement subset corresponding to 45,000 images for training. The ConvNeXt, Swin Transformer, and Vision Transformer models follow the same training procedure as for the DLMD. The Lensless Imaging Transformer is trained for 450,000 steps ($\approx 60$ epochs) using a learning rate of $3 \times 10^{-5}$.

\begin{backmatter}
\bmsection{Funding}
L.A.K. was supported by the National Science Foundation Graduate Research Fellowship Program under Grant DGE 2146752.
L.W. is a Chan Zuckerberg Biohub SF investigator. This work was supported by 
U.S. Air Force Office of Scientific Research Award No. FA955-22-1-0521.

\bmsection{Acknowledgment}
The authors thank Fanglin Linda Liu for phase mask manufacturing and Kristina Monakhova, Chaoying Gu, and Nick Antipa for helpful discussions. L.W. is a Chan Zuckerberg Biohub SF investigator.

\bmsection{Disclosures}
The authors declare no conflicts of interest.

\bmsection{Data Availability Statement}
All data and code are publicly available~\cite{ConvRML_Repo}.

\bmsection{Supplemental Document}
See Supplementary Material for additional results and implementation details.

\end{backmatter}

\bibliography{sample}

\clearpage

\renewcommand{\thesection}{S\arabic{section}}
\renewcommand{\thetable}{S\arabic{table}}
\renewcommand{\thefigure}{S\arabic{figure}}
\renewcommand\theequation{S\arabic{equation}}
\setcounter{figure}{0}
\setcounter{table}{0}
\setcounter{section}{0}
\setcounter{equation}{0}

\noindent
\textbf{\large ConvRML: high-quality lensless imaging with random multi-focal lenslets \\ Supplementary Material}

\vspace{0.1cm}

\noindent 
\textbf{Leyla~A.~Kabuli$^{1, \ast}$, Clara~S.~Hung$^{1}$, Vasilisa~Ponomarenko$^{1}$, Eric~Markley$^{1}$, Laura~Waller$^{1}$}

\noindent
\small$^1$Department of Electrical Engineering and Computer Sciences, University of California, Berkeley\\
\small$^\ast$Corresponding author. Email: lakabuli@berkeley.edu 
\\

\noindent 
This supplementary material contains additional results and implementation details.

\section{Additional reconstruction architecture evaluations}\label{s1}

\paragraph{Inference and training times}

We report inference times for a single image reconstruction in Table~\ref{tab:inference_times}. The inference time for ConvNeXt is shorter than both the large Swin Transformer and the Lensless Imaging Transformer~\cite{PanTransformer}. Inference times were calculated  by averaging 100 reconstruction passes on an NVIDIA RTX A6000 GPU after 10 warmup iterations, with GPU synchronization. As the Lensless Imaging Transformer reconstructs one channel at a time, its inference time for one RGB image corresponds to three sequential channel reconstructions.

\begin{table*}[h!tbp]
\centering
\caption{Single image reconstruction inference time for each model evaluated on the DiffuserCam Lensless Mirflickr Dataset (DLMD) and Parallel Lensless Dataset (PLD).}
\label{tab:inference_times}
\begin{tabular}{lcc}
\toprule
\textbf{Model}
& \textbf{DLMD inference (ms)} 
& \textbf{PLD inference (ms)} \\
\midrule 
U-Net        & 3.77 &   N/A \\
Vision Transformer & 3.42 & 3.46 \\
Swin Transformer & 105 & 125 \\
Lensless Imaging Transformer  & 128 & 150 \\
ConvNeXt & 55.3 & 61.3 \\

\bottomrule
\end{tabular}
\end{table*}

We report model training times and memory usage for the Parallel Lensless Dataset (PLD) in Table~\ref{tab:pld_train_time} and for the DiffuserCam Lensless Mirflickr Dataset (DLMD) in Table~\ref{tab:dlmd_train_time}, using the same number of training epochs for both datasets. ConvNeXt has training times that are shorter than the Lensless Imaging Transformer and the Swin Transformer for both datasets, while maintaining less memory usage than the Swin Transformer throughout. 

Reported results use the model checkpoint with the best validation loss within the training schedules specified in the main text. For the DLMD, this corresponds to step 210{,}000 ($\approx$ epoch 52) for the Lensless Imaging Transformer and the final checkpoint for all other models. For the PLD, this corresponds to epoch 34 for the Swin Transformer and the 100{,}000 measurement ConvNeXt, epoch 33 for the 75{,}000 measurement ConvNeXt, step 270{,}000 ($\approx$ epoch 36) for the Lensless Imaging Transformer, and the final checkpoint for all other models. Training and validation losses for all models decrease and then plateau, indicating stable training and convergence (Fig.~\ref{fig:losses}).

\begin{table*}
\centering
\caption{Training times and memory usage for all models trained on 50,000 RML measurements from the Parallel Lensless Dataset (PLD).}
\label{tab:pld_train_time}
\begin{tabular}{lccc}
\toprule
\textbf{Model}
& \textbf{Epochs}
& \textbf{Training time (hrs)} 
& \textbf{GPU memory} \\
\midrule

Vision Transformer & 70 & 6 & 1 GB \\
Swin Transformer & 35 & 537 & 37 GB  \\
Lensless Imaging Transformer & 60 & 190 & 26 GB  \\
ConvNeXt & 35 & 155 & 32 GB \\

\bottomrule
\end{tabular}
\end{table*}

\begin{table*}
\centering
\caption{Training times and memory usage for all models trained on the DiffuserCam Lensless Mirflickr Dataset (DLMD).}
\label{tab:dlmd_train_time}
\begin{tabular}{lcc}
\toprule
\textbf{Model}
& \textbf{Training time (hrs)} 
& \textbf{GPU memory} \\
\midrule

Vision Transformer & 3 & 1 GB \\
Swin Transformer & 259 & 33 GB  \\
Lensless Imaging Transformer  & 90 & 19 GB  \\
ConvNeXt & 72 & 29 GB \\ 

\bottomrule
\end{tabular}
\end{table*}

\paragraph{Training dataset size}

We investigate how training dataset size affects reconstruction performance using our 100,000 measurement Parallel Lensless Dataset (PLD). 
We study the tradeoffs between dataset size, reconstruction quality, and training cost by training the ConvNeXt architecture on subsets of 25,000, 50,000, 75,000, and all 100,000 of the RML measurements from the PLD.

\begin{figure}[hbtp]
\centering
    \includegraphics[width=0.95\textwidth]{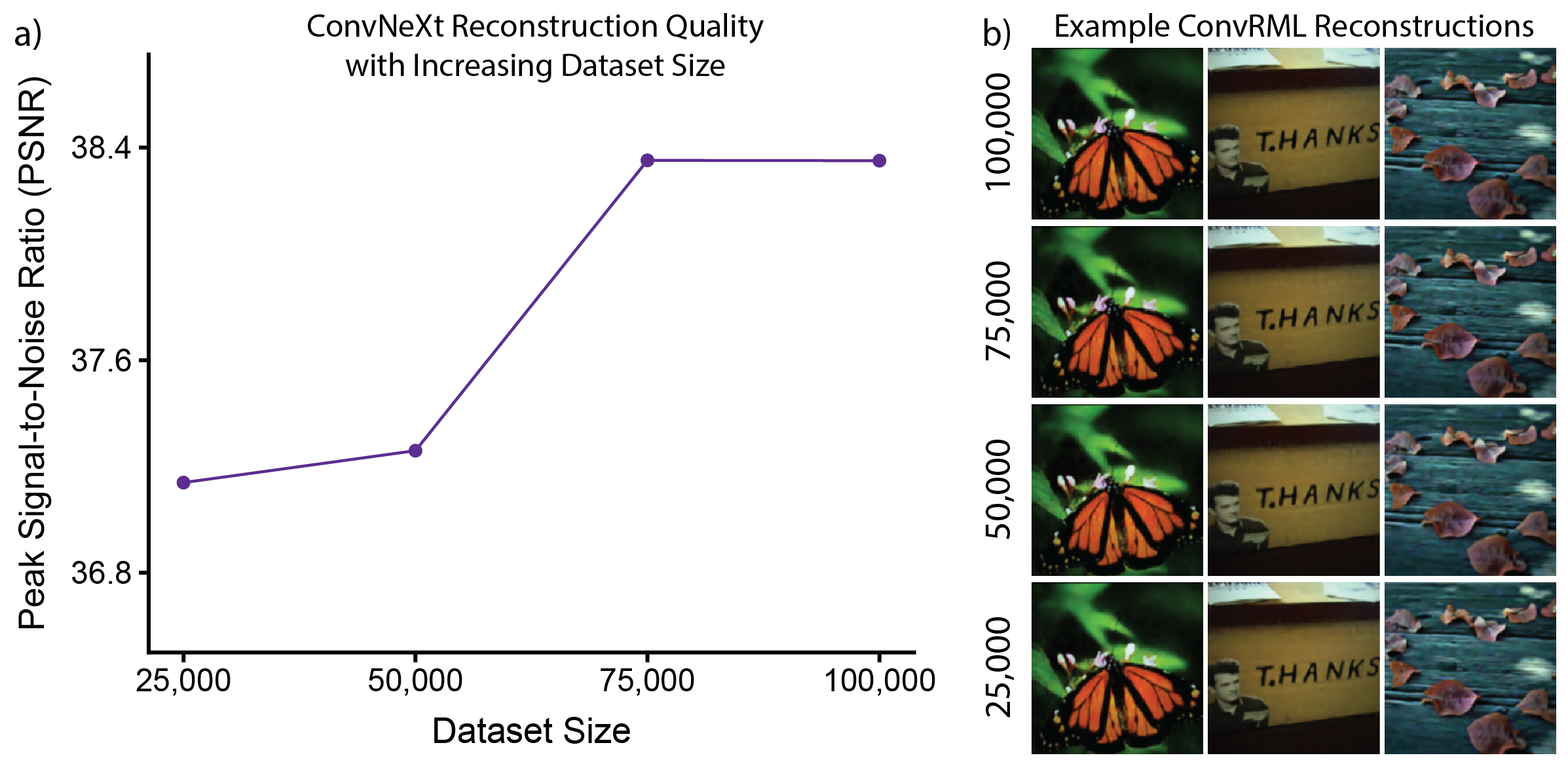}
    \caption{\textbf{Effect of training dataset size on ConvNeXt reconstruction quality for random multi-focal lenslet (RML) measurements from the Parallel Lensless Dataset.} a) Reconstruction peak signal-to-noise ratio (PSNR) as a function of training dataset size. b) Example ConvRML reconstructions for each training dataset size. Reconstruction quality improves as dataset size increases from 25,000 to 75,000 measurements and plateaus between 75,000 and 100,000 measurements. We use 50,000 measurements as our standardized dataset size to balance reconstruction quality and training cost across evaluated models.
    }
    \label{fig:dataset_size_sweep}
\end{figure}

Figure~\ref{fig:dataset_size_sweep}a reports reconstruction performance as a function of training dataset size. Reconstruction quality improves up to 75,000 measurements and then plateaus, while training time increases substantially with dataset size (Table~\ref{tab:dataset_sweep_train_time}). Although larger dataset sizes can provide additional quantitative improvement, reconstructions are visually similar across dataset sizes (Fig.~\ref{fig:dataset_size_sweep}b). We use 50,000 measurements as our standardized training dataset size for all relevant evaluations and comparisons to maintain high reconstruction quality while keeping training costs tractable across all evaluated architectures.

\begin{table*}[t]
\centering
\caption{ConvNeXt training times with increasing dataset size for random multi-focal lenslet (RML) measurements.}
\label{tab:dataset_sweep_train_time}
\begin{tabular}{lc}
\toprule
\textbf{Dataset size}
& \textbf{Training time (hrs)} \\
\midrule

25,000 & 69 \\
50,000 &  155 \\
75,000  & 234 \\
100,000 & 335 \\ 

\bottomrule
\end{tabular}
\end{table*}

\paragraph{Reconstruction quality with physics-based iterative reconstruction}

\begin{figure}[hbtp]
\centering
    \includegraphics[width=0.85\textwidth]{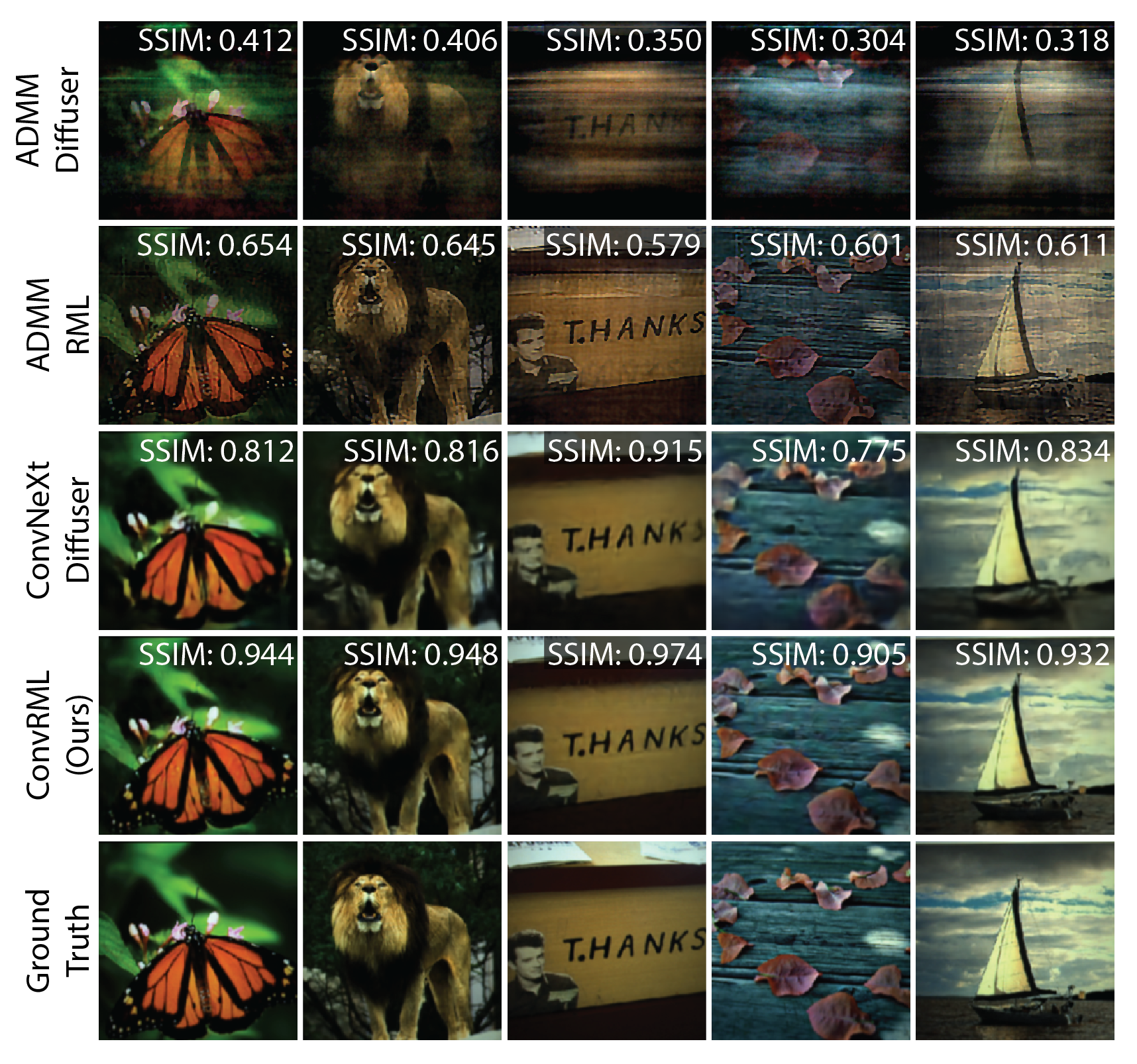}
    \caption{\textbf{Comparison of ConvNeXt and iterative alternating direction method of multipliers (ADMM) reconstruction.} ADMM reconstructions of diffuser and random multi-focal lenslet (RML) measurements have visible image artifacts and reduced image quality, while the ConvNeXt reconstructions for both diffuser and RML measurements closely match the ground truth images. Reconstruction quality is quantified by the structural similarity index measure (SSIM).}
    \label{fig:admmcomparison}
\end{figure}

In Fig.~\ref{fig:admmcomparison}, we compare reconstruction quality for our ConvNeXt model and alternating direction method of multipliers (ADMM)~\cite{ADMM} for diffuser and RML measurements.  ADMM is a representative physics-based iterative reconstruction baseline. We use 10 iterations of ADMM with parameters $\mathrm{\mu}_1 = 1e^{-4}$, $\mathrm{\mu}_2 = 1e^{-4}$, $\mathrm{\mu}_3 = 1e^{-4}$, $\tau = 2e^{-3}$, along with each system's point spread function (PSF). Each image is reconstructed in \SI{440}{\milli\second}. 

Using this fixed ADMM implementation, the RML reconstructions have higher structural similarity index measure (SSIM) than the diffuser reconstructions. This result demonstrates that the performance improvement contributed by the RML is not specific to the ConvNeXt architecture. ConvNeXt's pure deep learning-based approach provides significant quality improvement over ADMM's artifact-prone reconstructions for both diffuser and RML measurements, without requiring any knowledge of the system PSF or image formation model.

\section{Modulation transfer function evaluation}\label{s2}

\begin{figure}[hbtp]
\centering 
    \includegraphics[width=0.9\textwidth]{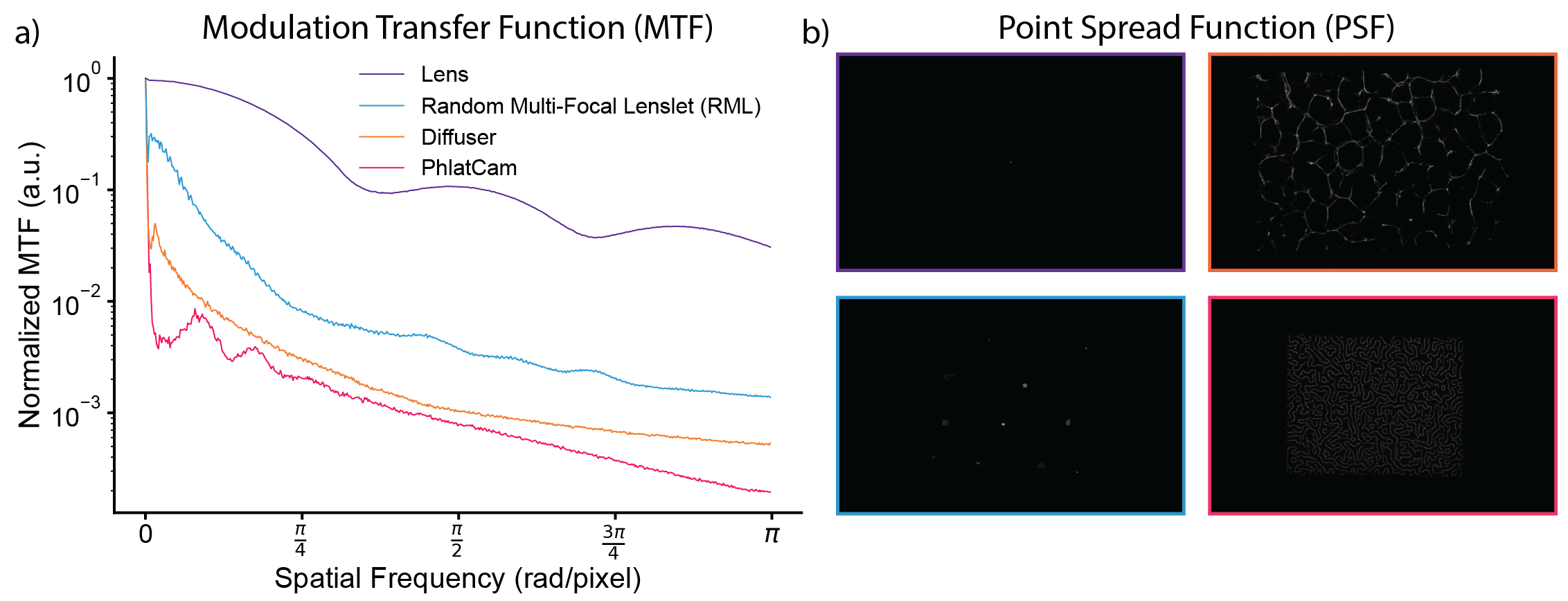}
    \caption{\textbf{Modulation transfer function (MTF) evaluation.} a) MTFs for random multi-focal lenslet (RML), diffuser, and PhlatCam point spread functions (PSFs) using the lens as an ideal reference. The RML has improved frequency transfer compared to both the diffuser and PhlatCam. 
    b) PSFs for each imaging system.
    }
    \label{fig:phlatcamMTF}
\end{figure}

In addition to the parallel system evaluation in the main text, we evaluate the frequency transfer abilities of our RML compared to the PhlatCam phase mask~\cite{PhlatCam}, a heuristically-designed mask with a high-multiplexing caustic pattern. In Fig.~\ref{fig:phlatcamMTF}, we compute the modulation transfer function (MTF) for each system's point spread function (PSF). Since the PhlatCam phase mask is not included in our parallel system setup, we use a publicly available PSF capture~\cite{PhlatCam}, which is cropped to match the aspect ratio of our sensors and rescaled to the same spatial dimensions prior to MTF computation.

The RML has superior frequency transfer compared to both PhlatCam and the diffuser. Notably, the diffuser PSF also outperforms the PhlatCam PSF, in contrast with prior MTF evaluations~\cite{Feng2022Learned, PhlatCam}. This finding highlights both the sensitivity of MTF analysis to experimental conditions and the variability inherent to Gaussian diffusers. These comparisons emphasize the importance of standardized evaluation and reproducible optical fabrication when comparing lensless imaging phase masks. 

\begin{figure}[hbtp]
\centering 
    \includegraphics[width=0.8\textwidth]{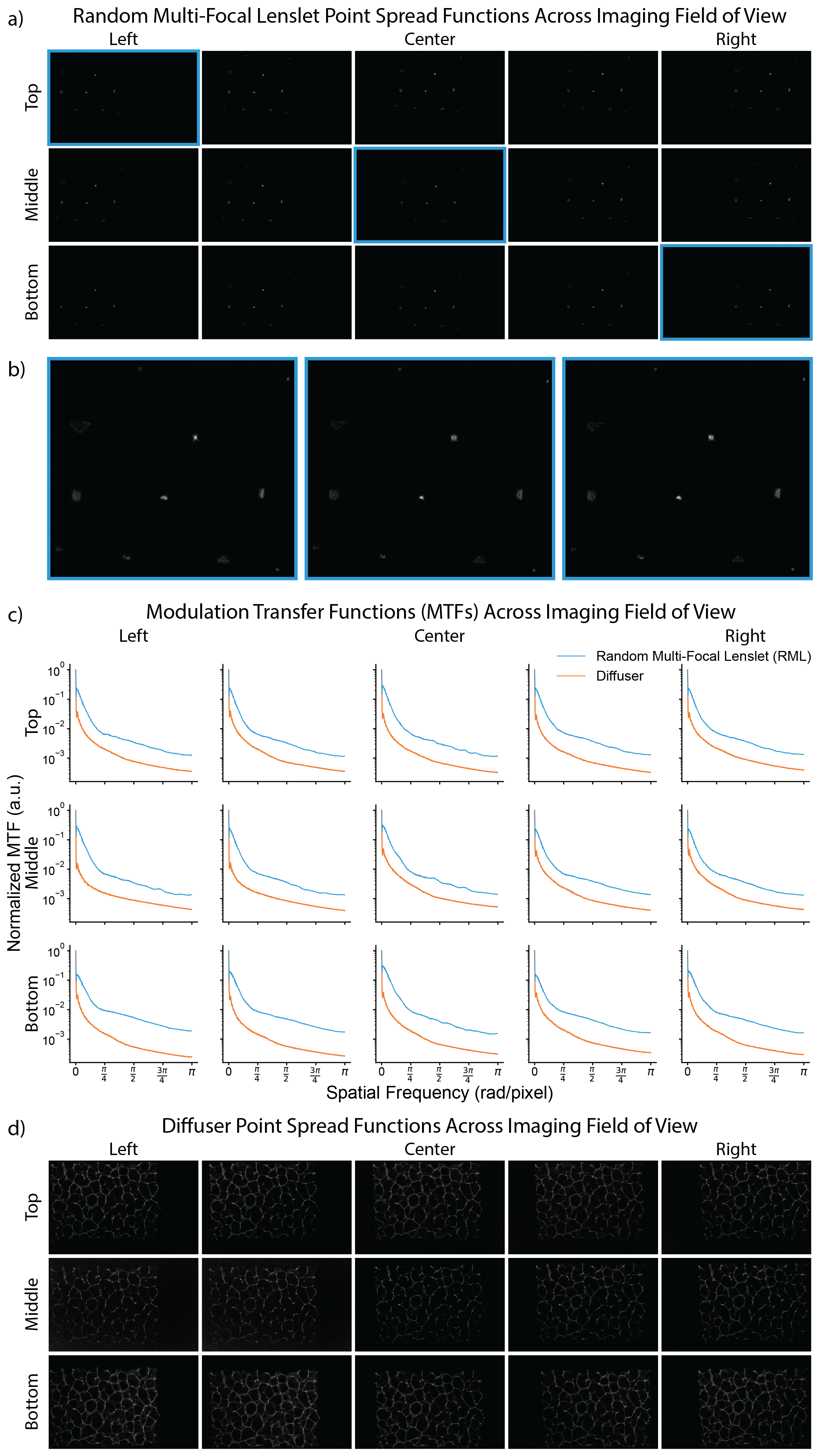}
    \caption{\textbf{Field-dependent point spread function (PSF) and modulation transfer function (MTF) evaluation.} a) Random multi-focal lenslet (RML) PSFs at 15 positions spanning the imaging field of view (FOV). The top left, center, and bottom right PSFs, outlined in blue, are registered to the center PSF and enlarged in b). RML PSFs have slight changes in focal spot shape and intensity across the FOV. c) MTFs for RML and diffuser PSFs at each corresponding field position. The RML maintains higher frequency transfer than the diffuser at all measured positions. d) Diffuser PSFs at 15 positions spanning the imaging FOV.
    }
    \label{fig:svpsfmtf}
\end{figure}

To characterize spatial variation, we capture experimental PSFs at 15 positions spanning the imaging field of view (FOV) by laterally translating a point source placed at the calibration plane for each imager (Fig.~\ref{fig:svpsfmtf}a, d). Enlarged RML PSFs show small changes in focal spot shape and relative intensity (Fig.~\ref{fig:svpsfmtf}b), suggesting that mild spatially varying effects are present. At each position, we compute MTFs for the corresponding PSFs (Fig.~\ref{fig:svpsfmtf}c). For both systems, the MTF profiles vary slightly across field positions, consistent with the PSF changes. The RML maintains higher frequency transfer than the diffuser at every measured position. 

\section{Real object reconstructions}\label{s3}
\begin{figure}[b]
\centering
    \includegraphics[width=0.95\textwidth]{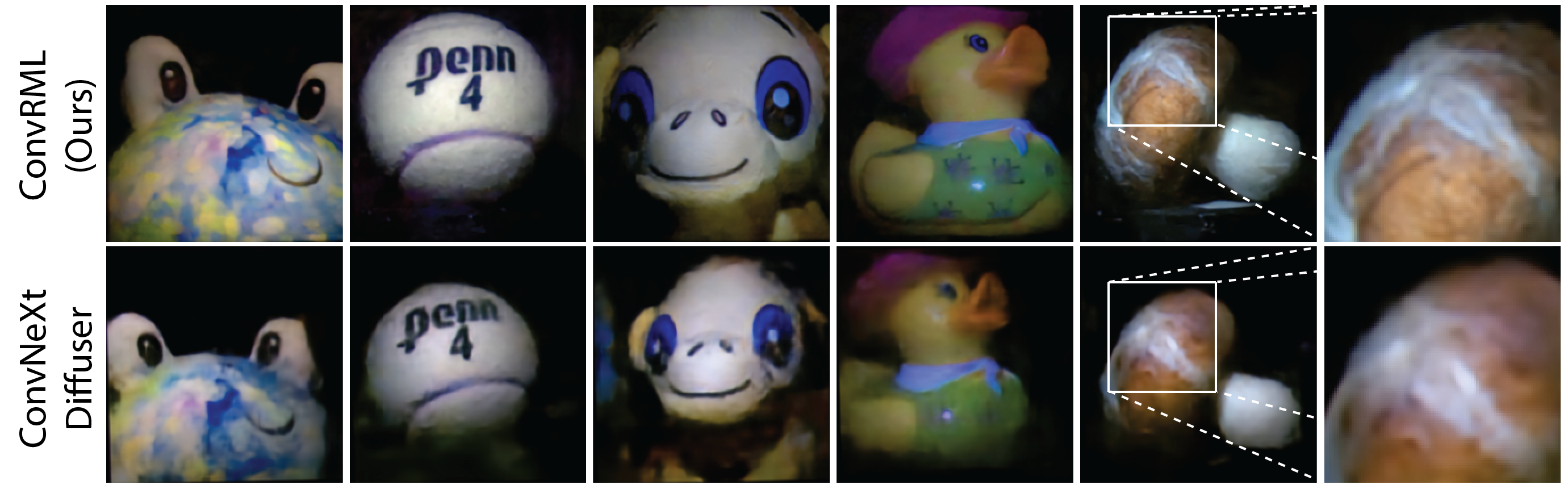}
    \caption{\textbf{Additional real object reconstructions.} 
    Reconstructions of real objects with varying surface properties using ConvNeXt models trained on displayed image datasets for the RML (top) and
    diffuser (bottom) lensless imaging systems.
    }
    \label{fig:supp_experimental}
\end{figure}

We provide additional reconstructions of real objects with varying materials and surface properties in Fig.~\ref{fig:supp_experimental}.
ConvRML recovers fine details and textures, whereas ConvNeXt reconstructions of diffuser measurements are blurred, as illustrated by the inset of the fibers on the felt mushroom. ConvRML also recovers the patterns on the duck torso while tolerating non-uniform illumination and specular effects, which introduce distortions in diffuser reconstructions. Minor differences in position and magnification are due to the practical limitations of manual object alignment across imaging systems. Real object reconstructions use ConvNeXt models trained on automatic white balance (AWB) datasets (Supplement Sec.~\ref{s5.3}), which match the AWB settings used for capturing these measurements.

\section{Random multi-focal lenslet phase mask}\label{s4}
\begin{figure}[hbtp]
\centering
    \includegraphics[width=0.6\textwidth]{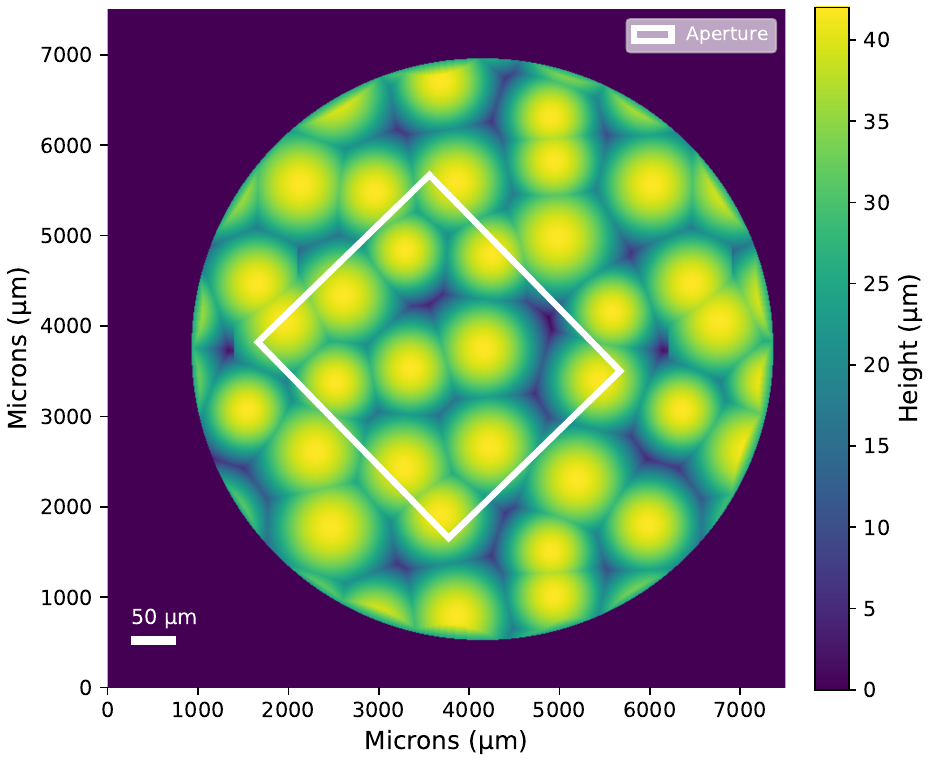}
    \caption{\textbf{Height map for generating the random multi-focal lenslet (RML) phase mask.} Surface height map for the RML design defined on a grid with \SI{10}{\micro\metre} pixel pitch. The white rectangle indicates the active region including the aperture placed on the fabricated optical element. 
    }
    \label{fig:heightmap}
\end{figure}

Figure~\ref{fig:heightmap} visualizes the height map used to generate the RML phase mask, with a white rectangle indicating the active aperture region (\SI{3.02}{\mm} by \SI{2.65}{\mm}) used for the RML system. 
For fabrication, the height map is defined on a $750 \times 750$ pixel surface grid with \SI{10}{\micro\metre} pixel pitch, corresponding to a total fabrication area of $7.5 \times 7.5$ \SI{}{\mm}. The fabrication file is provided with the PLD~\cite{ConvRML_Repo}.

Lenslet-based designs have been constrained by the lack of fabrication techniques capable of producing freeform optical elements with controlled surface variations. Two-photon lithography used for caustic phase mask fabrication~\cite{kcphasemask, PhlatCam} faced limitations in modulation depth, surface discretization, and large FOV stitching~\cite{Kyrollos2020Miniscope3D}, which restricted its utility for realizing smooth multi-focal lenslet designs. As a result, systems have relied on alternative, often manual, approaches that constrain the mask design space or limit precise surface control, limiting reproducibility and overall system quality. 
For example, microscopy-scale designs utilized multi-focal phase masks produced using mechanically-dented molds~\cite{LindaFourierDiffuserScope}, droplet curing~\cite{GraceDiffuserScope}, or hand-arranged uni-focal lenslets~\cite{Feng2025DeepInMiniscope}. 
Macro-scale designs consist of off-the-shelf regular lenslet arrays~\cite{xueCM2, xue2022CM2V2, SVFourierNet} as well as random uni-focal arrays made with mechanically-dented~\cite{NickRollingShutter} or 3D printed molds~\cite{Feng2021Geomscope, Feng2022Learned}.
 Our RML is distinguished from previous lenslet-based systems by the use of multi-focal lenslets for photography-scale imaging and precise, reproducible fabrication of our optical element.
Large-area grayscale lithography~\cite{nanoscribefabex} and freeform optical element manufacturing have resolved many of the fabrication constraints faced by earlier systems, providing the necessary capabilities to manufacture our multi-focal design.

\section{Parallel Lensless Dataset implementation}\label{s5}

\begin{figure}[t]
\centering
    \includegraphics[width=\textwidth]{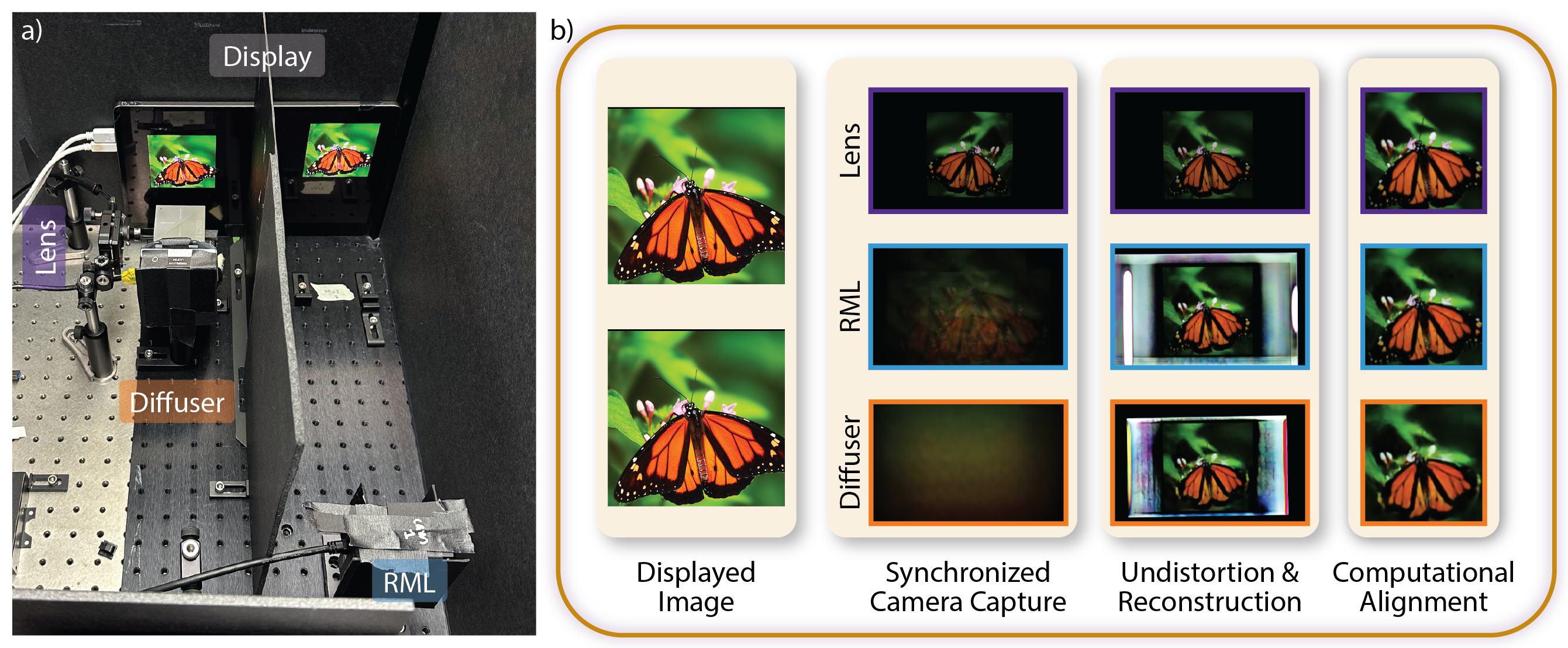}
    \caption{\textbf{Parallel dataset acquisition system and processing pipeline.}
    a) Experimental hardware setup with the RML and diffuser in parallel and a lens system incorporated with a beamsplitter. 
    b) Images from the MIRFLICKR-1M~\cite{mirflickr1M} dataset are displayed and captured synchronously by all three systems. Lens distortion and mirroring are corrected for the ground truth measurements from the lensed system, and RML and diffuser reconstructions are registered to the ground truth image and cropped to the center of the field-of-view.
    }
    \label{fig:dataset}
\end{figure}

For parallel system capture, the RML and diffuser systems are aligned to a display, with the lens system incorporated into the optical path of the diffuser with a beamsplitter (Fig.~\ref{fig:dataset}a).  We control for stray light using an enclosure placed around the parallel imaging system and non-reflective dividers between the RML and diffuser. 
Each sensor is mounted to standard 2-inch optical components with custom 3D-printed adapters. A step-by-step tutorial and computer-aided design files for the adapters are provided with the PLD~\cite{ConvRML_Repo}.

\paragraph{Automated dataset acquisition}\label{s5.1}
The Parallel Lensless Dataset (PLD) consists of the first 100,000 images of the MIRFLICKR-1M dataset~\cite{mirflickr1M}. We automatically display images with a Python acquisition script that uses the Pygame package~\cite{pygame}. 

Image display and data capture are synchronized across the three cameras with a \SI{200}{\milli\second} delay between each camera and a \SI{500}{\milli\second} delay between each image to ensure reliable image buffer saving. Sensor exposure times are calibrated using reference measurements to approximately match signal range across systems and avoid measurement saturation. Exposure times are set to \SI{80}{\milli\second}, \SI{25}{\milli\second}, and \SI{18}{\milli\second} for RML, diffuser, and lens systems, respectively. Captured measurements are saved directly to an external solid-state drive (SanDisk SDSSDE61-2T00-G25). Total acquisition time for the dataset is 40 hours. All system control scripts are released along with the PLD~\cite{ConvRML_Repo}.

\paragraph{Ground truth image registration} 
\label{s5.2}
Reconstructions from the RML and diffuser systems are computationally registered to ground truth images (Fig.~\ref{fig:dataset}b) for quantitative evaluation.
To correct for lens distortion in ground truth measurements, we calibrate the distortion profile using the OpenCV $11\times4$ asymmetric circle grid~\cite{opencv}. Ground truth measurements are mirrored across the $y$-axis to address the inversion introduced by the lens. 

For pixel-level alignment between lensless reconstructions and ground truth images, we estimate a transformation matrix (homography) from each lensless imager to the ground truth image (lens) space using Kornia~\cite{kornia}. This transformation corrects for shifts, rotations, and scale differences that cannot be fully corrected with manual optical alignment. Each homography is heuristically initialized and calibrated with $4 \times$ downsampled images reconstructed with 200 iterations of FISTA~\cite{FISTA}. For the diffuser, the homography is manually adjusted due to poor image quality in FISTA reconstructions. The color artifacts along the border of diffuser and RML reconstructions in Fig.~\ref{fig:dataset}b are introduced by the ConvNeXt-based reconstruction, which is trained on a cropped field-of-view. Regions outside this crop are not constrained during training.

\paragraph{Alternate automatic white balance dataset} \label{s5.3}

The PLD used in the main text consists of linear camera measurements captured without automatic white balance (AWB) or gamma correction. In addition to this primary dataset, we provide a variant of the dataset captured using typical camera processing settings, including AWB and gamma correction, which we refer to as AWB-PLD. The AWB-PLD extends our previously released 25{,}000-measurement AWB dataset~\cite{claraPWdataset} to 100{,}000 measurements. 

\begin{figure}
    \includegraphics[width=\textwidth]{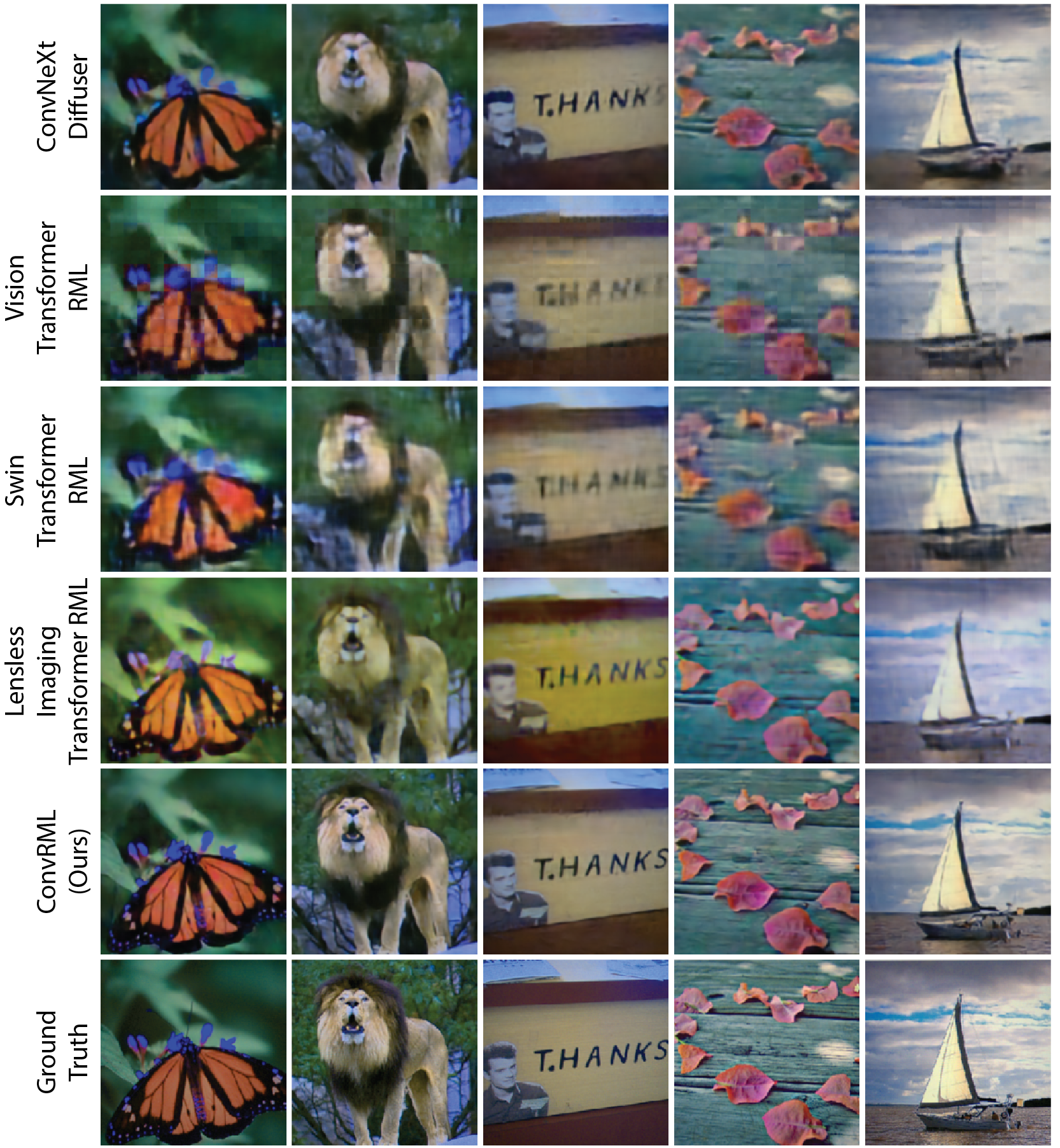}
    \caption{\textbf{Comparison of reconstruction models evaluated on our Automatic White Balance Parallel Lensless Dataset (AWB-PLD).} Our ConvRML system provides reconstructions that are best matched to ground truth images from the AWB-PLD. Although the AWB-PLD introduces image-dependent variations and a blue color cast on images, the ConvRML system maintains the highest reconstruction quality, consistent with the reconstruction results for the PLD in Fig.~5.}
    \label{fig:50k_algorithms_awb}
\end{figure}

\begin{figure}
    \includegraphics[width=\textwidth]{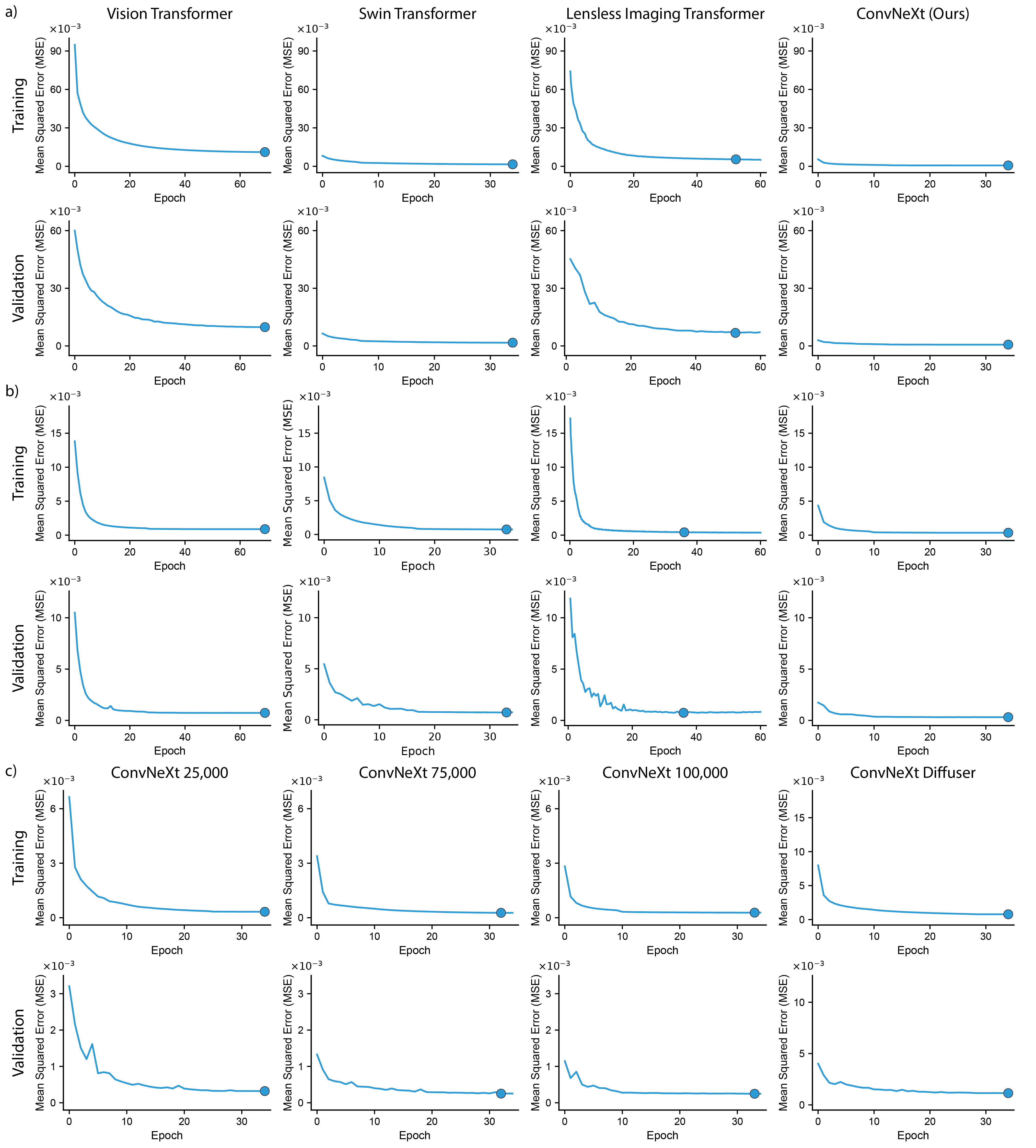}
    \caption{\textbf{Training and validation loss curves for evaluated models.} Training (top) and validation (bottom) mean squared error (MSE) loss curves for all models trained on a) the DiffuserCam Lensless Mirflickr Dataset (DLMD) and b) 50,000 random multi-focal lenslet (RML) measurements from the Parallel Lensless Dataset (PLD). c) Loss curves for ConvNeXt models trained on 25,000, 75,000, and 100,000 RML measurements from the PLD and 50,000 diffuser measurements from the PLD. Blue circles indicate the reported model checkpoints, selected based on the minimum validation loss. Across models and datasets, losses decrease and then plateau, indicating stable training.}
    \label{fig:losses}
\end{figure}

The camera processing in the AWB-PLD introduces image-dependent variations and channel scaling. In particular, this produces a blue color cast, which is especially noticeable in the butterfly and lion reconstructions in Fig.~\ref{fig:50k_algorithms_awb}. As a result, AWB-PLD images have reduced color fidelity compared to the original displayed MIRFLICKR images. For this reason, we focus on the PLD instead of the AWB-PLD. However, the AWB-PLD remains useful as a challenging reconstruction benchmark that contains additional color variation and nonlinear processing.

For completeness, we compare reconstruction architectures on the AWB-PLD in Fig.~\ref{fig:50k_algorithms_awb} and report metrics in Table~\ref{tab:rml_comparison_awb}. Although the absolute metrics are lower than those for the PLD (main text Table~2) due to the additional variability introduced by camera processing, the conclusions are consistent: the ConvRML system achieves the highest reconstruction quality, outperforming both the diffuser system and all attention-based architectures. The Lensless Imaging Transformer recovers fine details and structure, but has reduced metric values and strong color artifacts, consistent with its channel-wise reconstruction procedure~\cite{PanTransformer}. 

Real object measurements were also captured with AWB and gamma correction enabled. We reconstruct these measurements (main text Fig.~1, Supplement Fig.~\ref{fig:supp_experimental}) using ConvNeXt models trained on the AWB-PLD (Fig.~\ref{fig:50k_algorithms_awb}), matching the camera settings used for measurement capture. AWB-PLD models use the same architecture and training procedure as the PLD models, differing only in the training dataset used. 

\begin{table*}[h]
\centering
\caption{Reconstruction quality for models evaluated on 50,000 measurements from the Automatic White Balance Parallel Lensless Dataset (AWB-PLD).}
\label{tab:rml_comparison_awb}
\begin{tabular}{lccc} %
\toprule
\textbf{Model}
& \textbf{PSNR $\uparrow$}
& \textbf{SSIM $\uparrow$} 
& \textbf{LPIPS $\downarrow$}\\
\midrule
ConvNeXt Diffuser & 24.199 & 0.714  & 0.268 \\ %
Vision Transformer RML  & 23.500 & 0.673 & 0.329 \\ %
Swin Transformer RML & 23.201 & 0.674 & 0.361 \\ %
Lensless Imaging Transformer RML  & 20.506 & 0.674  & 0.352\\ %
ConvRML (Ours)       & \textbf{27.187} & \textbf{0.830}  & \textbf{0.126} \\ %

\bottomrule
\end{tabular}
\end{table*}

\clearpage 

\end{document}